\documentclass[aps, pra, superscriptaddress, floatfix, reprint]{revtex4-2}
\setcitestyle{super,open={},close={}}

\usepackage{hyperref, graphicx, xspace}
\hypersetup{hidelinks, allcolors=blue, colorlinks=true}

\usepackage{multirow,booktabs,siunitx,natmove,threeparttable,enumitem}

\usepackage{float}
\makeatletter
\let\newfloat\newfloat@ltx
\makeatother

\AtBeginDocument{\RenewCommandCopy\qty\SI}
\DeclareSIUnit{\angstrom}{\textup{\AA}}
\DeclareSIUnit{\hartrees}{hartrees}

\usepackage{algorithm, algpseudocode}
\newcommand*{\PySCF}{\textsc{PySCF}\xspace}
\newcommand*{\PySCFforge}{\textsc{PySCF-forge}\xspace}
\newcommand*{\geomeTRIC}{\textsc{geomeTRIC}\xspace}
\newcommand*{\libxc}{\textsc{libxc}\xspace}
\newcommand*{\libcint}{\textsc{libcint}\xspace}
\newcommand*{\ASE}{\textsc{ASE}\xspace}

\usepackage[version=4]{mhchem}
\usepackage{cleveref}

\usepackage{physics, calrsfs, mathtools, bm}
\DeclareMathAlphabet{\pazocal}{OMS}{zplm}{m}{n}

\begin{document}

\title{Weighted Active Space Protocol for Multireference Machine-Learned Potentials}

\author{Aniruddha Seal}\thanks{These authors contributed equally to this work.}
\affiliation{Department of Chemistry and Chicago Center for Theoretical Chemistry, University of Chicago, Chicago, IL 60637, USA}

\author{Simone Perego}\thanks{These authors contributed equally to this work.}
\affiliation{Atomistic Simulations, Italian Institute of Technology, 16156 Genova, Italy}

\author{Matthew R. Hennefarth}
\affiliation{Department of Chemistry and Chicago Center for Theoretical Chemistry, University of Chicago, Chicago, IL 60637, USA}

\author{Umberto Raucci}
\affiliation{Atomistic Simulations, Italian Institute of Technology, 16156 Genova, Italy}

\author{Luigi Bonati}
\affiliation{Atomistic Simulations, Italian Institute of Technology, 16156 Genova, Italy}

\author{Andrew L. Ferguson} 
\affiliation{Department of Chemistry and Chicago Center for Theoretical Chemistry, University of Chicago, Chicago, IL 60637, USA}
\affiliation{Pritzker School of Molecular Engineering, University of Chicago, Chicago, IL 60637, USA}

\author{Michele Parrinello} \email[corresponding author: ]{michele.parrinello@iit.it}
\affiliation{Atomistic Simulations, Italian Institute of Technology, 16156 Genova, Italy}

\author{Laura Gagliardi} \email[corresponding author: ]{lgagliardi@uchicago.edu} 
\affiliation{Department of Chemistry and Chicago Center for Theoretical Chemistry, University of Chicago, Chicago, IL 60637, USA}
\affiliation{Pritzker School of Molecular Engineering, University of Chicago, Chicago, IL 60637, USA}

\date{\today}

\begin{abstract}

Multireference methods such as multiconfiguration pair-density functional theory (MC-PDFT) offer an effective means of capturing electronic correlation in systems with significant multiconfigurational character. However, their application to train machine learning-based interatomic potentials (MLPs) for catalytic dynamics has been challenging due to the sensitivity of multireference calculations to the underlying active space, which complicates achieving consistent energies and gradients across diverse nuclear configurations. To overcome this limitation, we introduce the Weighted Active-Space Protocol (WASP), a systematic approach to assign a consistent active space for a given system across uncorrelated configurations. By integrating WASP with MLPs and enhanced sampling techniques, we propose a data-efficient active learning cycle that enables the training of an MLP on multireference data. We demonstrate the method on the \ce{TiC+}-catalyzed \ce{C-H} activation of methane, a reaction that poses challenges for Kohn-Sham density functional theory due to its significant multireference character. This framework enables accurate and efficient modeling of catalytic dynamics, establishing a new paradigm for simulating complex reactive processes beyond the limits of conventional electronic-structure methods.

\end{abstract}

\maketitle

\newpage
\section*{Introduction}

Machine learning-based interatomic potentials (MLPs) have emerged as powerful tools for approximating the potential energy surface by learning the mapping from nuclear configurations to energies and forces based on high-level quantum mechanical data. Once trained, MLPs can reproduce quantum mechanical accuracy at a fraction of the computational cost, enabling simulations of larger systems and longer timescales than would be feasible with \textit{ab initio} methods \cite{deringer_machine_2019,schran_machine_2021,unke_machine_2021}. These models have shown great promise in applications ranging from catalyst dynamics under \textit{operando} conditions \cite{bonati_role_2023,yang2023reactant,perego2024dynamics,omranpour_machine_2025} to phase transitions \cite{bonati2018silicon, jinnouchi_phase_2019,deringer_origins_2021,abou2024unraveling}, and excited state dynamics \cite{dral2021molecular,raucci2025capturing,westermayr2020machine}. Nevertheless, the accuracy and reliability of MLPs depend critically on the availability of high-quality quantum mechanical calculations \cite{kulichenko_data_2024}.
 
While Kohn-Sham density functional theory (KS-DFT) has proven to be a reliable and cost-effective source of training data for organic reactions and biomolecular processes \cite{christensen_orbnet_2021,wang_ab_2024}, it often fails for systems and processes with significant multiconfigurational character. These systems typically include open-shell transition metals with multiple low-lying electronic states, electronically excited states, and bond-breaking process. Accurate treatment of multiconfigurational systems requires a multireference electronic-structure methods. The complete active space self-consistent field (CASSCF) \cite{roos_complete_1980,roos_complete_1987} method is a widely successful multireference method that can capture static electron correlation yielding a qualitatively accurate wave function, but it often fails to capture all of the dynamic electron correlation especially outside of the active space. To regain quantitative accuracy, CASSCF is often followed by post-SCF approaches to account for dynamic correlation outside of the active space, with common choices including multiconfiguration pair-density functional theory 
(MC-PDFT) \cite{li_manni_multiconfiguration_2014,sharma_multiconfiguration_2021,zhou_electronic_2022}, $n$-electron valence state second-order perturbation theory (NEVPT2) \cite{angeli_introduction_2001}, and complete active space second-order perturbation theory (CASPT2) \cite{andersson_second-order_1990}.

Among these post-SCF methods, MC-PDFT offers a good compromise between accuracy and computational efficiency. It computes the energy of a multiconfigurational wave function using a functional form inspired by KS-DFT:
\begin{equation}
    E^{\mathrm{PDFT}} = V^{\mathrm{nuc}} + \sum_{pq} h^q_p\gamma^p_q + \frac{1}{2}\sum_{pqrs} g^{qs}_{pr}\gamma^p_q\gamma^r_s + E^{\mathrm{ot}}[\rho,\Pi]
\end{equation}
Here, $V^\mathrm{nuc}$ is the nuclear-nuclear repulsion energy, $h^q_p$ and $g^{qs}_{pr}$ are the one- and two-electron integrals, and $\gamma^p_q$ is the one-electron reduced density matrix. The key ingredient distinguishing MC-PDFT from KS-DFT is the on-top energy functional, $E^{\mathrm{ot}}\bqty{\rho,\Pi}$, which depends on both the electron density ($\rho$) and the on-top pair density ($\Pi$), and captures correlation effects beyond the active space. For a more detailed description, we refer the reader to \citet{zhou_electronic_2022} and \citet{sharma_multiconfiguration_2021}.

MC-PDFT has been shown to produce vertical excitation energies \cite{hoyer_multiconfiguration_2016,king_large-scale_2022} and reaction barriers \cite{carlson_multiconfiguration_2015, wardzala_organic_2024,mitchell_application_2022,maity_role_2022} of similar quality to the more computationally expensive NEVPT2 and CASPT2. It has also recently been used to study the nonadiabatic dynamics of organic molecules \cite{calio_nonadiabatic_2022,hennefarth_semiclassical_2024}, further demonstrating its versatility. However, its application to long-timescale bond-breaking dynamics in transition metal systems remains largely unexplored due to the practical limitations of CASSCF. Although MC-PDFT supports analytical gradients\cite{scottAnalytic2021,ScottAnalytic2020,SandAnalytic2018} and can in principle be used for \textit{ab initio} molecular dynamics (AIMD), such simulations become computationally infeasible when spanning broad regions of nuclear configuration space, especially in the context of reactive events that require extensive sampling.

To alleviate this bottleneck, MLPs trained with MC-PDFT data offer a promising solution that enable efficient exploration of multireference-quality potential energy surfaces. However, constructing reactive MLPs based on multireference data presents two major challenges. The first is due to the challenges of constructing a reactive MLP, which requires collecting a diverse and representative training set that captures configurations along the entire reaction pathway including uncorrelated geometries near the transition state region~\cite{yang2022using,david2025arcann}. This challenge is increasingly addressed through a combination of enhanced sampling methods such as metadynamics \cite{laio_escaping_2002}, on-the-fly probability enhanced sampling (OPES) \cite{invernizzi_rethinking_2020,trizio2024advanced}, and active learning strategies \cite{smith_less_2018,ang_active_2021,bonati_role_2023,yang2022using,herr_metadynamics_2018,benayad_prebiotic_2024} that iteratively expand the training set by using model uncertainty to identify and label the most relevant configurations for improving model accuracy. Recently, we introduced a data-efficient active learning (DEAL) \cite{perego_data_2024} protocol that enables uniformly accurate reactive modeling with fewer \textit{ab initio} calculations.

The second challenge, which is specific to active-space based multireference methods, is ensuring a consistent active-space selection across a diverse set of uncorrelated geometries. MC-PDFT properties are uniquely determined by the underlying CASSCF wave function. Unlike in AIMD or nudged elastic band calculations where molecular properties can be smoothly propagated between adjacent steps, active learning strategies involve single-point calculations on geometries that may be widely separated in nuclear configuration space. Because the CASSCF wave function optimization is highly sensitive to the initial active-space guess and can converge to local minima for different geometries for which there does not exist a continuous path to connect the two solutions, it complicates the definition of a stable and transferable active space across the training set \cite{saade_excited_2024}. This makes active-space selection non-trivial, as distinct local minima may not be adiabatically connected across the nuclear configuration space. This challenge becomes even more critical in transition metal systems that often require large active spaces to accurately capture open-shell character and strong multiconfigurational effects. In such cases, it is particularly easy to converge to different electronic solutions, leading to discontinuities in the underlying electronic potential energy that ultimately prevents the training of reliable MLPs.

Although a number of automated active-space selection strategies have been developed, relying on natural orbital occupations \cite{khedkar_active_2019}, atomic valence rules \cite{sayfutyarova_automated_2017}, orbital entanglement metrics \cite{stein_autocas_2019,king_ranked-orbital_2021}, or machine learning \cite{jeong_automation_2020}, these approaches are typically tailored for optimized equilibrium structures. Extensions to reaction pathways have utilized orbital-pair selection schemes and frame-based initialization \cite{wen_strategies_2024, han_complete_2020}. In practice, these methods are primarily applied to equilibrium geometries; and when applied to non-equilibrium geometries, such as those sampled during dynamics, they often fail to provide consistent and adiabatically connected active spaces that limit their applicability in an active learning workflow.

In this paper, we introduce the Weighted Active Space Protocol (WASP), which ensures consistent active-space assignment throughout the MLP active learning cycle. By providing a systematic and uniform definition of active spaces across geometries, WASP enables consistent labelling of active-space multireference calculations for the training datasets. Combined with DEAL, an enhanced sampling-based active learning scheme, this framework enables the construction of MLPs trained directly on MC-PDFT energies and gradients -- a first-of-its-kind demonstration of multireference-quality MLPs. The methodology presented here is general and applicable to other post-CASSCF methods such as CASPT2 and NEVPT2, opening the door to efficient simulations of strongly correlated systems.

\begin{figure*}
  \includegraphics[width=\textwidth]{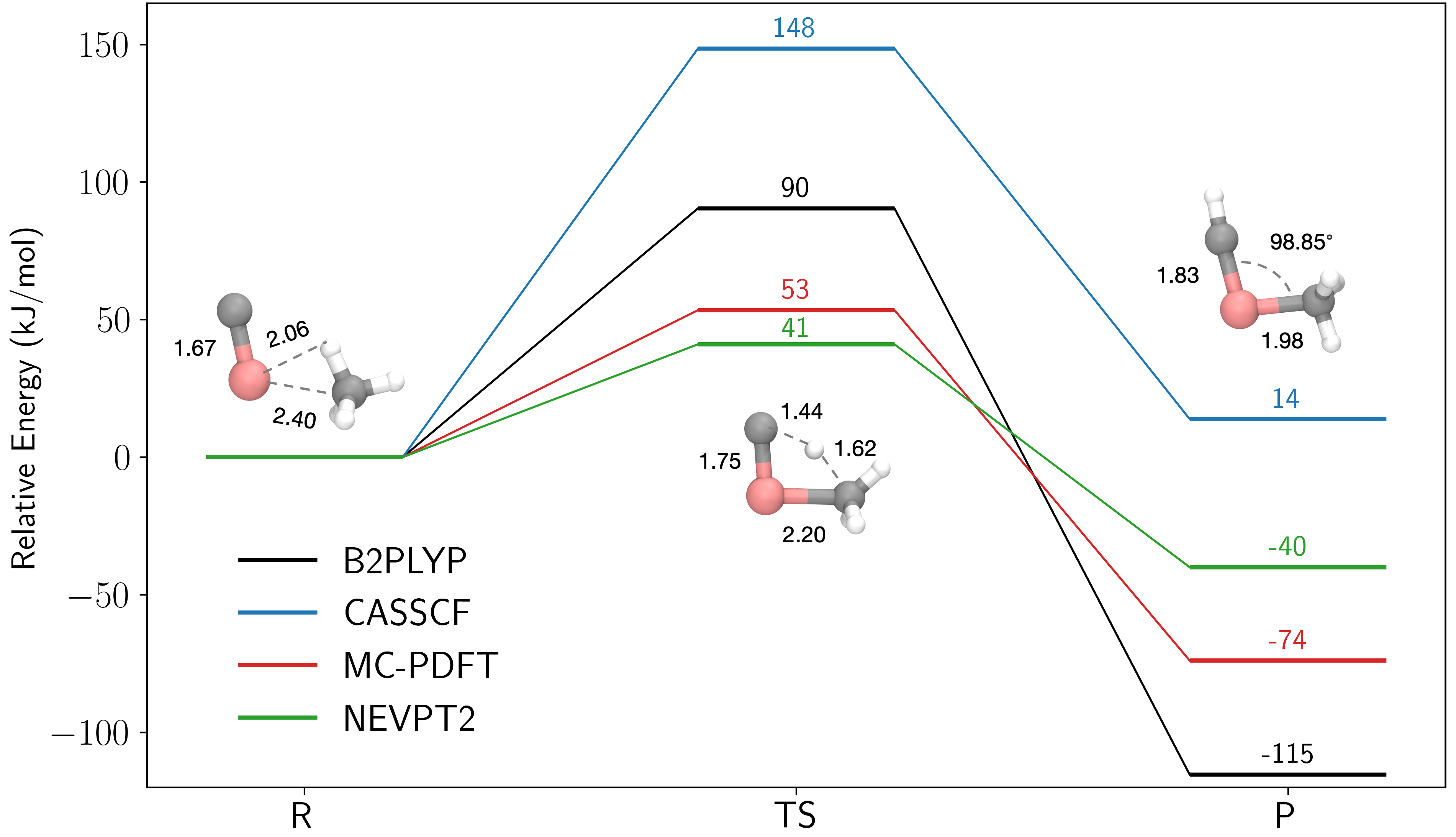}
  \caption{\label{fig:pes} Potential energy profiles for the initial proton-coupled electron transfer step of the $^{2}$\ce{[TiC+]}-catalyzed \ce{C-H} activation of methane ($\mathbf{R} \rightarrow \mathbf{TS} \rightarrow \mathbf{P}$) computed using state-averaged CASSCF over the lowest two doublet states with a (7e, 9o) active space, MC-PDFT with the tPBE on-top functional (using the SA-CASSCF as the reference wave function), NEVPT2 (data from \citet{geng_intrinsic_2019}), and KS-DFT with B2PLYP functional. Geometrical features of the ground-state structures ($\mathbf{R}$ and $\mathbf{P}$ optimized with MC-PDFT and $\mathbf{TS}$ from \citet{geng_intrinsic_2019}) are marked with the bond lengths (in \unit{\angstrom}) and bond angles (in degrees).}
\end{figure*}

We validate our approach on the rate-determining step of \ce{TiC+}-catalyzed \ce{C-H} activation of methane, a prototypical case where KS-DFT fails to provide an accurate description \cite{geng_intrinsic_2019}. The reaction proceeds via a proton-coupled electron transfer in which the doublet ground-state \ce{TiC+} approaches methane to form an encounter complex ($\mathbf{R}$), followed by hydrogen atom migration through a four-membered transition state ($\mathbf{TS}$), leading to the formation of a new \ce{C-H} bond in the intermediate ($\mathbf{P}$) (\cref{fig:pes}). We modeled the reaction using a 7 electron in 9 orbital (7e,9o) active space as used by \citet{geng_intrinsic_2019}. As shown in \cref{fig:pes}, the calculated barrier height using MC-PDFT (with the tPBE on-top functional) is \qty{53}{\kilo\joule\per\mole}, in agreement with the prior NEVPT2 result \cite{geng_intrinsic_2019}. In contrast, the CASSCF method significantly overestimates the barrier at \qty{148}{\kilo\joule\per\mole}. DFT calculations using the B2PLYP functional give a barrier of \qty{90}{\kilo\joule\per\mole} that appears implausibly high given that the reaction is known to proceed experimentally in the gas phase at room temperature \cite{geng_intrinsic_2019}. These comparisons underscore both the limitations of KS-DFT and the accuracy of MC-PDFT, making it an ideally suited multireference method for studying this system. This reaction serves as a stringent and representative test case for evaluating the ability of our protocol to generate consistent active spaces and train a MLP on high-quality multireference data.

\section*{Results and discussion}
\subsection*{WASP: Weighted active-space protocol}

Multireference electronic-structure methods like MC-PDFT depend on an optimized CASSCF wave function, defined by its molecular orbital (MO) coefficients and configuration interaction (CI) expansion to yield energies and forces that are uniquely determined by that wave function. In practice, the optimization is highly sensitive to the initial guess for the MOs. Even with a fixed active-space size, different choices of initial orbitals or active space composition can converge to different local minima. Since training a MLP requires a single continuous potential energy surface, it is crucial to assign each geometry an active-space guess that ensures consistency across hundreds of uncorrelated geometries.

To address this, we introduce the weighted active space protocol (WASP). The key idea of WASP is to interpolate MO coefficient matrices from a library of wave function guesses to construct high-quality, geometry-specific initial guesses. Given a new geometry, $\vb{G}$, and a set of previously visited structures $\Bqty{\vb{G}_\alpha}$ with corresponding MO coefficient matrices $\Bqty{\vb{C}_\alpha}$, WASP first selects geometries $\Bqty{\vb{G}_\beta}$ in a $\delta$ neighborhood such that $d\pqty{\vb{G}, \vb{G}_\beta} \leq \delta$. We use the root mean square deviation (RMSD) as our metric $d$ to select the neighboring geometries, although other metric choices are possible. Next, it computes a weighted sum of the MO coefficients $\Bqty{\vb{C}_\beta}$ as
\begin{equation}
    \vb{C} = \sum_{\beta} \tilde{w}_\beta \vb{C}_\beta
\end{equation}
where the weights are defined as:
\begin{equation}
    w_\beta = \frac{1}{d\pqty{\vb{G}, \vb{G}_\beta}}
\end{equation}
\begin{equation}
    \tilde{w}_\beta = \frac{w_\beta}{\sum_\beta w_\beta}
\end{equation}
The weights, $w_\beta$, are defined such that they approach $1$ as the new geometry approaches any previously visited geometry ($d\pqty{\vb{G}, \vb{G}_\beta} \rightarrow 0$ then $\tilde{w}_\beta \rightarrow 1$) and $0$ as the distance grows ($d\pqty{\vb{G}, \vb{G}_\beta} \rightarrow \infty$ then $\tilde{w}_\beta \rightarrow 0$). The normalized weights, $\tilde{w}_\beta$, are defined so that there sum is equal to $1$. If $\vb{G}$ is the same as some geometry within the library, that is there is some $\vb{G}_\beta$ such that $d\pqty{\vb{G}, \vb{G}_\beta} = 0$, then we just take $\vb{C}_\beta$ to be $\vb{C}$. Finally, the resulting guess coefficient matrix $\vb{C}$ is orthonormalized using the overlap matrix of atomic orbitals $\vb{S}\pqty{\vb{G}}$ via a standard procedure (such as Gram-Schmidt orthonormalization procedure) to yield the final orthonormalized MO guess $\vb{Q}$. The complete workflow is outlined in \cref{alg:wasp}.

\begin{algorithm*}
\caption{Weighted Active Space Protocol (WASP)}
\begin{algorithmic}[1]

\Statex \textbf{Input:} Geometries $\Bqty{\vb{G}_\alpha}$ and their corresponding MO coefficient matrices $\Bqty{\vb{C}_\alpha}$, new geometry $\vb{G}$

\State Select neighbors ($\Bqty{\vb{G}_\beta}$) from $\Bqty{\vb{G}_\alpha}$ such that $d(\vb{G}, \vb{G}_\beta) \leq \delta$ 

\State Compute weights $\tilde{w}_\beta$ given geometries $\vb{G}_\beta$
\begin{equation*}
    w_\beta = \frac{1}{d\pqty{\vb{G}, \vb{G}_\beta}}, \quad \tilde{w}_\beta = \frac{w_\beta}{\sum_{\beta} w_\beta}
\end{equation*}

\State Compute guess MO coefficient matrix $\vb{C}$ for $\vb{G}$
\begin{equation*}
    \vb{C} = \sum_{\beta} \tilde{w}_\beta \vb{C}_\beta
\end{equation*}

\State Orthonormalize $\vb{C}$ under $\vb{S}\pqty{\vb{G}}$
\begin{equation*}
    \vb{C} \xrightarrow{\vb{S}\pqty{\vb{G}}} \vb{Q}
\end{equation*}

\Statex \textbf{Output:} Guess MO coefficient matrix $\vb{Q}$ for geometry $\vb{G}$

\end{algorithmic}
\label{alg:wasp}
\end{algorithm*}

\begin{figure*}
  \centering
  \includegraphics[width=\linewidth]{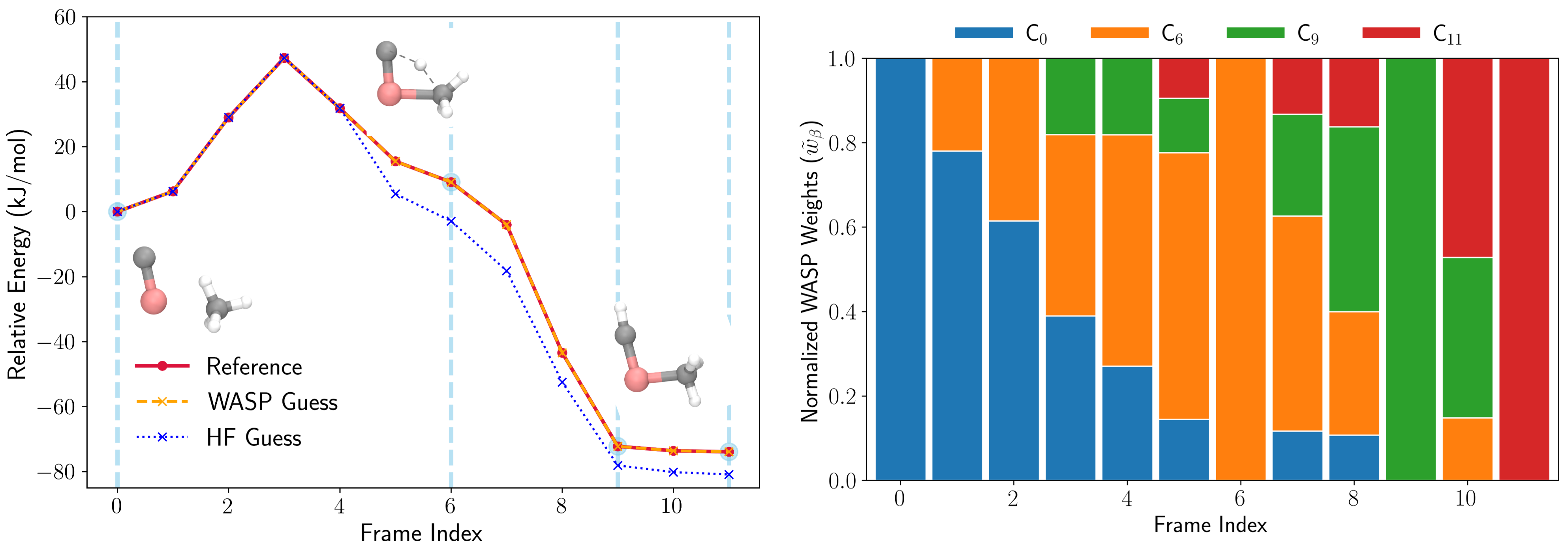}
  \caption{\label{fig:wasp} Demonstration of WASP on a reaction pathway connecting $\mathbf{R}$ and $\mathbf{P}$. (Left) Comparison of single-point MC-PDFT energies obtained using WASP active-space guesses and Hartree–Fock (HF) guesses, benchmarked against reference MC-PDFT energies. The reference MC-PDFT energy for each geometry was computed using a converged CASSCF wave function as the initial guess from each subsequent geometry, with the orbitals for the starting geometry selected as shown in fig.~S1. (Right) Normalized weights of MO coefficients ($\tilde{w}_\beta$) computed in WASP for each geometry from using $\Bqty{\vb{G}_\alpha}$, illustrating how the algorithm assigns initial orbital guesses to geometries along the reaction pathway.}
\end{figure*}

We validated WASP on a representative reaction pathway connecting $\mathbf{R}$ and $\mathbf{P}$. In Section SII of the Supporting Information, we describe how the checkpoint library of just four geometry-MO coefficient pairs $\Bqty{\vb{G}_\alpha, \vb{C}_\alpha}$ out of the twelve frames were selected and the heuristics behind choosing the cutoff parameter $\delta =$ \qty{0.8}{\angstrom}. We then compared MC-PDFT single-point energies obtained using WASP-generated orbital guesses to those obtained with a reference protocol, where each geometry was initialized using the converged wave function from the previous image. \Cref{fig:wasp}a shows that the WASP protocol was able to reproduce the reference MC-PDFT energies with high fidelity across the entire pathway. Furthermore, analysis of the assigned weights shown in \cref{fig:wasp}b confirmed that WASP generates smoothly varying orbital guesses, consistent with the adiabatic evolution of the wave function. In contrast, MC-PDFT energies based on Hartree–Fock (HF) guesses showed significant deviations, particularly beyond the transition state. These results establish WASP as a reliable and efficient orbital initialization scheme for use in multireference electronic-structure based machine learning workflows.

\subsection*{Active learning of multireference machine-learned potentials}

\begin{figure*}
  \includegraphics[width=\textwidth]{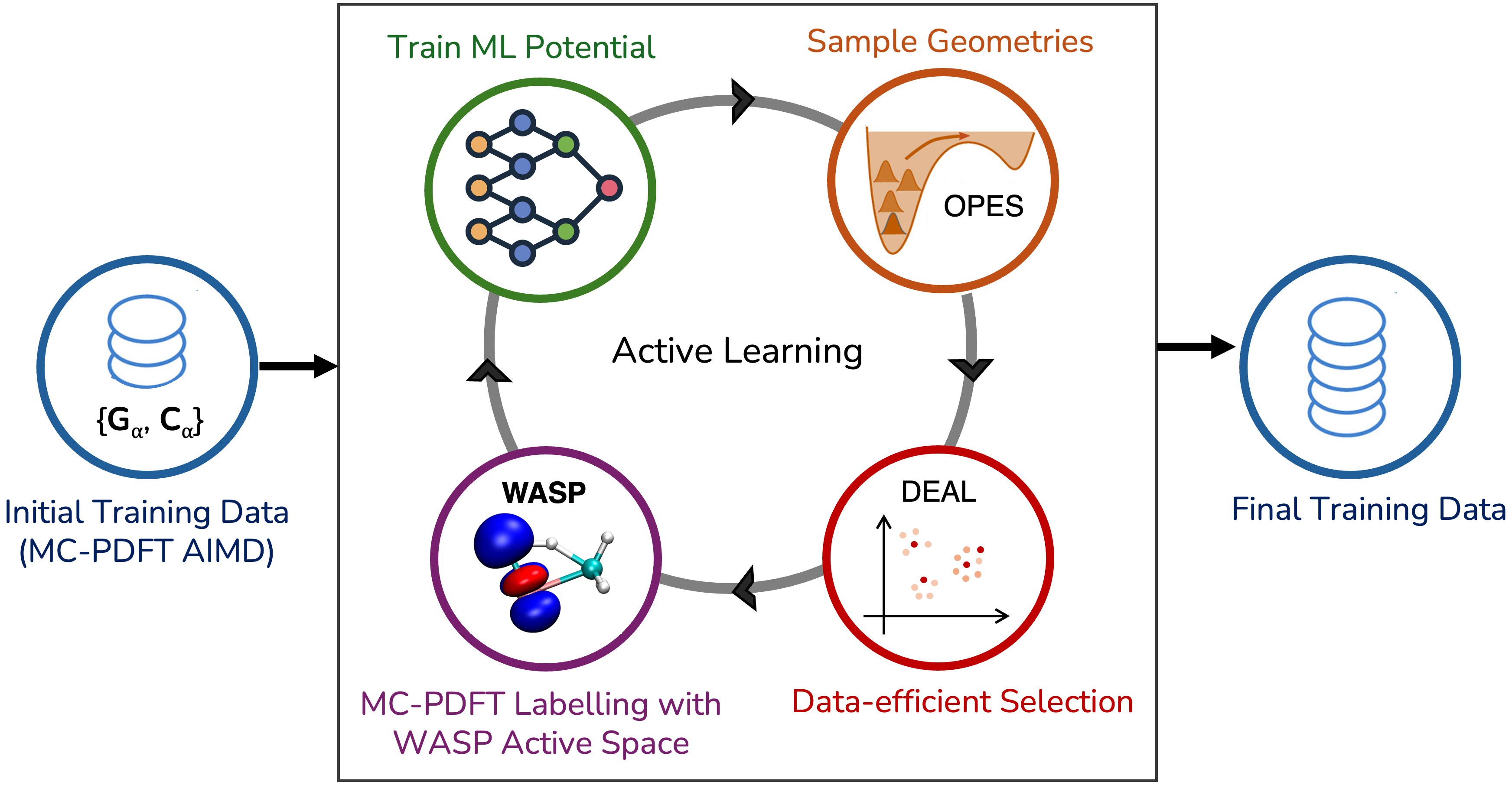}
  \caption{\label{fig:al_schematic} Active learning workflow to train multireference machine-learned potentials with MC-PDFT.}
\end{figure*}
The integration of MLPs with multireference electronic-structure methods, such as MC-PDFT, requires a data-efficient and automated workflow that ensures incremental sampling of relevant configurations and consistency in data labelling. To this end, we propose an active learning framework that combines data-efficient active learning schemes based on enhanced sampling with robust wave function guess initialization through WASP. The complete loop is designed to generate high-quality training data while ensuring that MC-PDFT energy and forces remain smoothly connected across the sampled configuration space. The steps of the workflow, as illustrated in \cref{fig:al_schematic}, are outlined below:

\begin{enumerate}[leftmargin=*]
    \item \textbf{Initial training set construction} \\
    We begin by generating a small ensemble of reference geometries, $\bqty{\vb{G}_\alpha}$, through Born–Oppenheimer AIMD integrated with MC-PDFT analytical gradients \cite{SandAnalytic2018, ScottAnalytic2020}. From these trajectories, we extract the MC-PDFT energies $\Bqty{E_{\alpha}}$, forces $\Bqty{\vb{f}_{\alpha}}$, and the MO coefficient matrices $\Bqty{\vb{C}_\alpha}$. The data collected populate both the initial training database for the MLP and the WASP checkpoint library.
    
    \item \textbf{Machine-learned potential training} \\
    We train a MLP on this initial dataset. In particular, we use MACE \cite{Batatia2022MACE:Fields}, an equivariant graph neural network that is data-efficient \cite{Batatia2022MACE:Fields,perego_data_2024} and well-suited for capturing the complex interactions typical of transition metal systems \cite{owen_complexity_2024}. Although we employ MACE here, WASP is fully compatible with alternative MLP training architectures.
    
    \item \textbf{(Enhanced) Sampling of new geometries} \\
    The trained MLP is then used to perform molecular dynamic simulations, systematically exploring relevant regions of nuclear configuration space and generating new candidate geometries for labelling. To build the dataset incrementally, we first run unbiased simulations to sample each metastable basin. Once the trained potential gives sufficiently reliable energies and forces in each basin, we switch to OPES enhanced sampling \cite{invernizzi_rethinking_2020} to capture reactive transitions and harvest the corresponding structures.
    
    \item \textbf{Data-efficient selection} \\
    From the ensemble of sampled geometries, the data-efficient active learning (DEAL) framework \cite{perego_data_2024} identifies a minimal number of configurations that should be labeled with MC-PDFT data. DEAL first applies a query-by-committee strategy to select structures with high model uncertainty, and then it uses the uncertainty of a Gaussian process to extract a minimal dataset of non-redundant geometries (see Materials and Methods). Since only structures that are structurally different from those already collected previously are selected, they represent ideal candidates not only for the expansion of the training set, but also for the WASP library.
    
    \item \textbf{MC-PDFT labelling with WASP guess} \\
    For each selected geometry, WASP constructs an initial multiconfigurational wave function guess by interpolating MO coefficients from nearby entries in the checkpoint library. These interpolated orbitals then seed CASSCF optimizations that yield MC-PDFT single‐point energies and gradients to label the geometries.
    
    \item \textbf{Incremental training set update} \\
    The MC-PDFT labelling and update of the WASP checkpoint library is performed incrementally on the candidates from step 5 to ensure smooth and consistent active-space selection throughout the sampling process. To prioritize well-supported regions of configuration space, we begin by labelling configurations that are closest to the configurations in the library, as measured by their minimum RMSD. To further limit extrapolation, we introduce an energy-based filter that excludes configurations whose predicted energy (from the MLP) significantly deviates from the corresponding MC-PDFT value (see Materials and Methods for details). This dual-filter strategy reinforces the reliability of the data labelling process and ensures an incremental update of the dataset limited to the regions where the active-space interpolation and MLP generalization remains accurate.

    \item \textbf{Iterate} \\
     The MACE model is then retrained on the expanded dataset, and the cycle of sampling, selection, labelling, and retraining continues (steps 2 through 6) until the model’s accuracy along the reaction pathway converges and the (biased) molecular dynamic simulations are stable. The converged model is subsequently deployed for production simulations.
\end{enumerate}

\subsection*{Machine-learned potential for \ce{TiC+}-catalyzed methane activation}

We applied our active learning protocol to construct a machine learning potential for the \ce{TiC+}-catalyzed C–H activation of methane. To provide robust electronic structure references, we employed a state-averaged CASSCF (SA-CASSCF) wave function throughout the active learning process \cite{saade_excited_2024, marie_excited_2023}.

\begin{figure}
    \centering
    \includegraphics[width=\columnwidth]{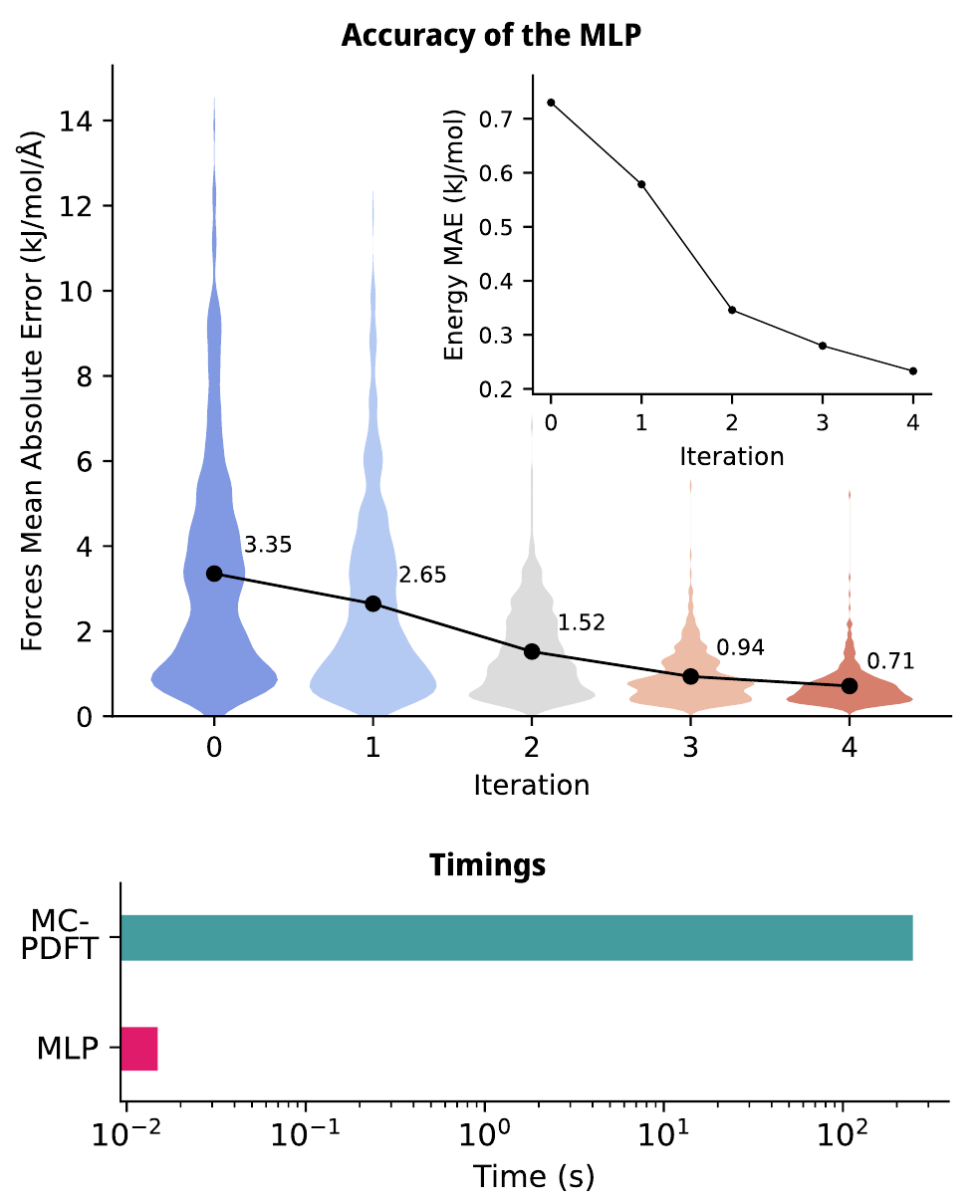}
    \caption{(Top) Violin plot of forces mean absolute error (MAE) on test set for the MACE potential across successive active-learning iterations. For forces, each violin illustrates the distribution of the MAE per geometry within a given iteration. The solid black line marks the overall average over the full test set. The inset displays the energy MAE for the same set across each active-learning round. (Bottom) Comparison of wall time for one gradient calculation of the MACE potential (on a NVIDIA-A100 GPU) trained on the MC-PDFT data and analytical MC-PDFT gradients (on two cores with 8 GB of memory on an AMD EPYC processor)}
    \label{fig:MLP_error}
\end{figure}

Initially, we constructed the reference library from six independent Born–Oppenheimer AIMD simulations, each \qty{0.5}{ps} long (1000 frames), three initialized at the reactant basin and three at the product basin with different velocity distributions. To reduce redundancy arising from the strong correlation among these configurations, we used the DEAL algorithm to extract a representative subset of 229 configurations, which served as our initial training dataset (see Materials and Methods for details).

We then trained an initial MLP and applied the above-described active learning scheme over four iterations. In each cycle, we performed both equilibrium simulations in the reactant and product basins, as well as reactive biased simulations—except during the first cycle, where only equilibrium simulations were included. Approximately 150 new configurations were added per cycle. The final MLP was trained on a dataset comprising 831 configurations. To visualize the expanding coverage of the configurational space, a convex hull was constructed from the configurations gathered in each round (fig.~S5), illustrating how the MLP progressively explores new regions of the reaction landscape.

\begin{figure*}
    \centering
    \includegraphics[width=1.\linewidth]{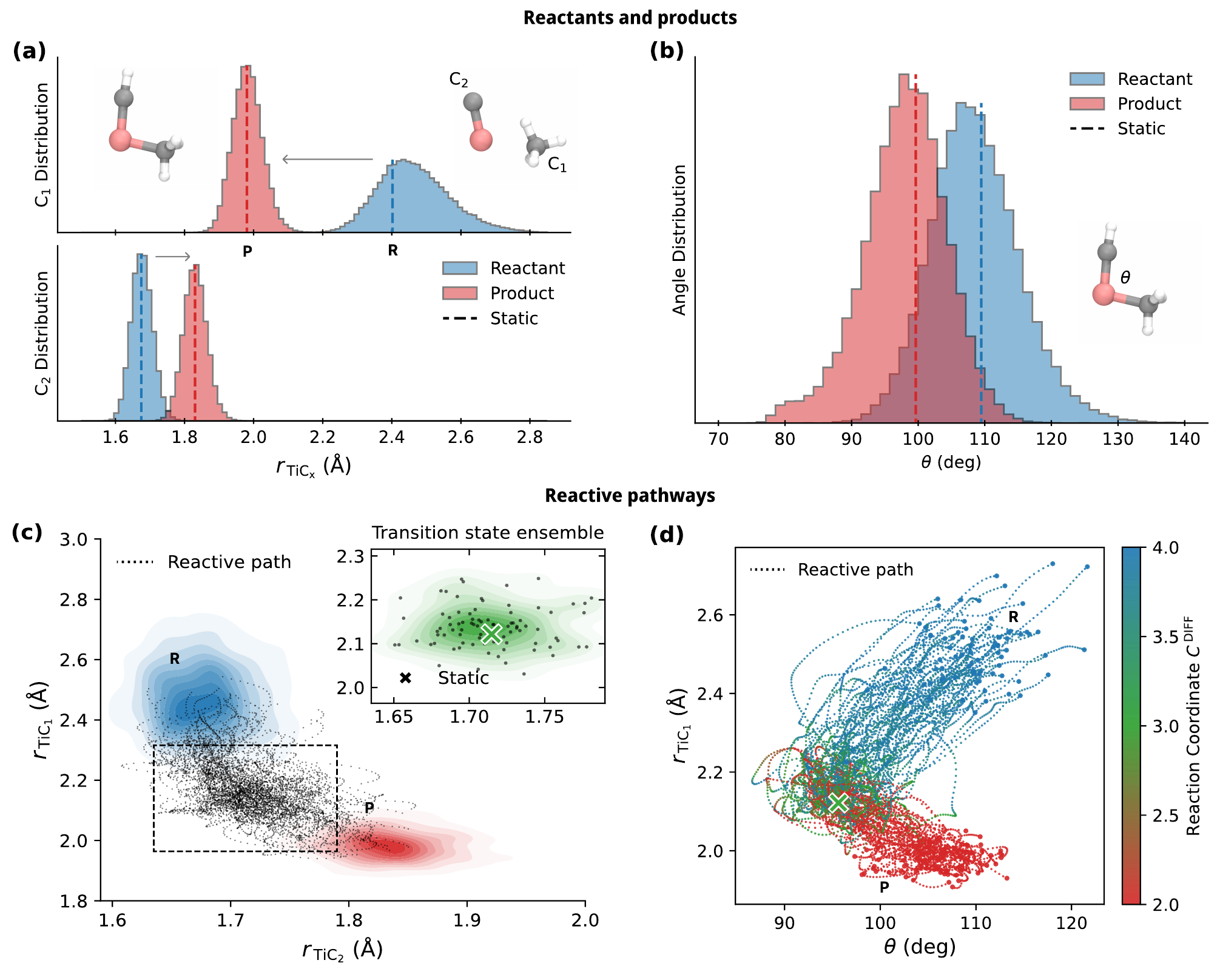}
    \caption{\label{fig:MLP_results} Dynamics analysis using a machine-learned potential trained on MC-PDFT data. (a) Distribution of \ce{Ti-C} distances ($r_{\ce{TiC_x}}$ and (b) \ce{C-Ti-C} bond angles ($\theta$) from 1\,ns-long equilibrium molecular dynamics simulations at \qty{300}{\kelvin} at the reactant ($\mathbf{R}$, blue) and product ($\mathbf{P}$, red) geometries. (c) Projection of reactive OPES flooding trajectories onto the $r_{\ce{TiC1}}$ and $r_{\ce{TiC1}}$ coordinates. Each trajectory is reported for \qty{20}{\pico\second} around the transition state (TS), which we identified as the frame for which the reaction coordinate $C^{\text{DIFF}}$ reaches the values of 2.9. This is equivalent to the value of $C^{\text{DIFF}}$ in the static TS.  The inset highlights the distribution of the transition-state (TS) ensemble. (d) Projection of the same trajectories onto the $r_{\ce{TiC1}}$ and $\theta$ internal coordinates colored by the difference in coordination number between the two carbon atoms. The static values are obtained from the \qty{0}{\kelvin} optimized reaction pathway reported in fig.~S9. }
\end{figure*}

In \cref{fig:MLP_error} we report the evolution of the accuracy of the potential as a function of iterations, validated on test dataset. We define the force error for any particular geometry to be the mean absolute error (MAE) between the MC-PDFT force and the MLP force. In particular, we calculated the distribution of force errors, which we reported together with their average value. At the beginning, the distribution is very broad, with peaks much higher than the mean (up to 4 times). The iterative procedure significantly lowers the mean value, but especially the tails of the distribution. In fact, in the end, we obtain a distribution with an average of only \qty{0.7}{\kilo\joule\per\mole\per\angstrom} for the forces. Similarly, the MAE on the energies decreases until it reaches \qty{0.35}{\kilo\joule\per\mole} (\cref{fig:MLP_error}). These results confirm that the protocol enables the construction of a highly accurate, multireference-quality MLP for strongly correlated transition metal systems. They also provide confirmation of the reliability of the WASP orbital-guess protocol and the library expansion scheme, as without consistent labelling of the energies and forces, we could not have learned them with this high accuracy. Furthermore, the MACE model achieves prediction timescales of \qty{\approx 15}{\ms} per timestep, compared to approximately four minutes for a MC-PDFT gradient calculation, thus drastically lowering the computational barrier to studying complex catalytic dynamics. 

\subsection*{MC-PDFT Quality Reaction Dynamics Using Machine Learning Potential}

Using the trained model, we initiated our study of reactive dynamics by sampling the equilibrium conformational space of both reactant and product states. For each, we ran \qty{1}{\ns} molecular dynamics simulations at \qty{300}{\kelvin}. From these trajectories, we extracted structural descriptors such as the \ce{Ti-C} bond distances ($r_{\ce{TiC}}$) and \ce{C-Ti-C} bond angle ($\theta$) (\cref{fig:MLP_results}a-b).
 
The \ce{Ti-C} distance distributions reveal that in the reactant state, methane is loosely bound to \ce{TiC+}, with an average \ce{Ti-C1} distance of \qty{2.4}{\angstrom} (\cref{fig:MLP_results}a). In contrast, after the hydrogen transfer, the $r_{\ce{TiC1}}$ distance contracts to about \qty{2.0}{\angstrom}, reflecting the formation of the \ce{Ti-C1} bond. Moreover, the \ce{Ti-C2} bond slightly elongates due to the redistribution of electronic density. 
$\theta$ shifts from approximately \qty{108}{\degree} in the reactant toward \qty{98}{\degree} at the product state, further corroborating the structural rearrangement (\cref{fig:MLP_results}b). The same behavior is observed for the \ce{C-C} distance, which is reported in fig.~S7.

To directly observe reactive events and extract statistics on the mechanistic and kinetic aspects of the reaction, we performed 100 independent molecular dynamic simulations using the OPES flooding approach, where a bias potential is applied only in the reactants' basin and leaves the transition‐state region unperturbed. By projecting the ensemble of reactive trajectories onto key geometric variables, we identified the sequence of structural changes that underpin the reaction mechanism (\cref{fig:MLP_results}c-d, fig.~S8). Mapping the reactive trajectories onto the two-dimensional coordinate space defined by the $r_{\ce{TiC1}}$ and $r_{\ce{TiC2}}$ bond distances, we observe a concerted reaction pathway (\cref{fig:MLP_results}c).
Notably, the broad spread of these trajectories reveals an ensemble of transition‐state geometries at finite temperature, highlighted in the inset of \cref{fig:MLP_results}c. We identified these geometries as the frame where the reaction coordinate $C^{\text{DIFF}}$ reaches the values of 2.9, consistent with its value for the optimized TS. 

This ensemble spans a broad range of \ce{Ti-C} distances, indicating a certain degree of variability in bonding character in the transition state region. This stands in contrast to the traditional picture of a single, well-defined transition state and underscores the importance of sampling-based approaches in capturing the full reactivity landscape at finite temperature.

Further insight can be obtained upon mapping the reactive pathways in the $\theta$ and $r_{\ce{TiC1}}$ coordinate space and tracking the difference in coordination number between the \ce{TiC+} carbon and the methane carbon ($C^\mathrm{diff}$) (which decreases from four in the reactant complex, to three at the transition state, and finally to two in the product), we can follow the reaction dynamics underlying the transition (\cref{fig:MLP_results}d). The reaction unfolds as the electron-deficient \ce{TiC+} interacts with methane, leading to the weakening of one of the \ce{C-H} bonds. This is evidenced by a \ce{C-H} bond starting to vibrate at a higher frequency during the \textbf{R}$\rightarrow$\textbf{TS} transition, as shown in Movie S1.
In addition, fig.~S8 shows that $\theta$ transiently decreases toward \qty{90}{\degree}, bringing the two carbon atoms closer together. This geometric shift facilitates electron transfer from the methyl group to \ce{TiC+}, while proton transfer simultaneously occurs from the methane carbon to the carbon on \ce{TiC+}. Following this key step, the $\theta$ angle increases and the system relaxes into the product geometry.

Finally, to quantify the kinetics of the $\mathbf{R}\rightarrow\mathbf{P}$ transition, we computed the cumulative distribution of transition times across all reactive events obtaining a mean first passage time equal to \qty{135}{\micro\second} (fig.~S10). Assuming transition state theory and applying the Eyring equation\cite{eyring_activated_1935} with a transmission coefficient of 1, this timescale corresponds to a free energy barrier of roughly \qty{50}{\kilo\joule\per\mole}. However, this direct estimate of the transition times from the molecular dynamics simulations is far more powerful, since it is based on statistical sampling of rare events and requires no further assumptions. Obtaining such extensive sampling of reaction dynamics lies beyond the practical capabilities of conventional DFT methods and is entirely out of reach for multireference approaches. The combination of machine learning potentials and enhanced sampling techniques enables access to experimentally relevant timescales and offers detailed insight into both the mechanism and kinetics of complex reactions.

\section*{Conclusion}
In this manuscript, we have presented a robust framework for training accurate machine learning-based interatomic potentials based on multireference electronic structure data. This approach overcomes the key limitation in integrating multireference electronic-structure methods with data-driven modeling: the consistent labelling of the energy and forces across diverse uncorrelated nuclear configurations. This challenge is especially critical in reactions, where the electronic structure can vary dramatically along the reaction pathway. To address this, we introduce a novel molecular orbital guessing scheme, WASP, which allows to maintain a stable active space across the training set. 

The general design of WASP makes it broadly applicable to other active-space based multireference approaches, including but not limited to CASSCF, CASPT2, and NEVPT2, where robust initial orbital guesses are equally essential. Although demonstrated here with MC-PDFT and MACE, the methodology can be readily adapted to alternative machine learning architectures and integrated into post-SCF workflows.

Looking ahead, the WASP-MLP framework is particularly promising for studying larger and more complex catalytic systems, including those with bulky ligands or significant conformational flexibility around the metal center. In such systems, accessing multireference-quality free energy surfaces is essential to accurately capture the reactivity, dynamics, and entropy-driven effects.

Moreover, the methodology opens exciting directions for further development. Transfer learning strategies could be employed to leverage pretrained representations and further improve data-efficiency \cite{franken} as well as accelerate model training by reusing the learned representations on a cheaper level of theory \cite{chen2023data,bocus2025operando}.
Additionally, this framework offers a systematic approach to benchmark multireference methods for catalytic reaction dynamics. 

Finally, while the mixing of the active-space initial guess in WASP currently relies on RMSD-based metrics, future extensions could incorporate chemically informed distance measures such as those based on ACE descriptors. This would allow more nuanced weighting schemes that better reflect underlying electronic and structural similarities, further improving the reliability and generalization of the active-space interpolation process.

Together, these developments establish the WASP-MLP paradigm as a powerful and flexible toolkit for exploring reactivity in strongly correlated, high-dimensional chemical systems with unprecedented accuracy and efficiency.

\section*{Materials and methods}

\subsection*{Electronic structure calculations}

All MC-PDFT and SA-CASSCF calculations were performed using \PySCF \cite{sunPySCF2018, sunRecent2020} (version 2.4.0, commit \texttt{0bc590ffddec9}), compiled with \libcint \cite{SunLibcint2015} (version 6.0.0), \libxc \cite{MarquesLibxc2012, LehtolaRecent2018} (version 6.1.0), and \PySCFforge \cite{PySCFforge2025} (version 1.0.0, commit \texttt{1077e48d5efab407}), an extension module for PySCF. Geometry optimizations were carried out using the \geomeTRIC plugin \cite{wangGeometry2016} (version 1.0) interfaced with \PySCF. All calculations used the def2-TZVP basis set \cite{def2tzvp}, and state averaging was performed over the lowest two doublet states. The tPBE \cite{li_manni_multiconfiguration_2014} on-top functional was employed for all MC-PDFT calculations with a numerical quadrature grid size of 6. No spatial symmetry constraints were applied. AIMD with MC-PDFT were performed using \ASE \cite{ase_2017} (version 3.23.0b1, commit \texttt{28a0a1f1988e3}), employing the Velocity Verlet integrator (NVE ensemble) with a timestep of \qty{0.5}{\fs}. Density fitting was used to accelerate the MC-PDFT calculations \cite{scottAnalytic2021}. Initial velocities were assigned from a Maxwell-Boltzmann distribution corresponding to \qty{300}{\kelvin}.

\subsection*{Machine-learned potential training details}
We employ the MACE\cite{Batatia2022MACE:Fields} (v0.3.7) graph neural network architecture to construct machine-learned interatomic potentials. All the models used in this work are equivariant, with $L=1$ and 128 channels, and use a radial cutoff of \qty{6}{\angstrom}. The training sets consist of configurations selected by the DEAL active learning framework and validated through a consistency check (see below). The validation and test sets are composed of $0.5\times$ the number of configurations obtained from Gaussian selection, randomly sampled among those preselected by query-by-committee but not selected, and subsequently passing the consistency check. The models are optimized using the AMSGrad optimizer, minimizing a weighted root mean square error of energies and forces. Training is carried out in two stages for a total of 400 epochs. In the first stage, the models are trained for 300 epochs using a learning rate of 0.01 and a batch size of 8, with energy and force loss weights set to 1 and 100, respectively. In the second stage, the best models from the initial training are further refined to improve energy predictions, using a learning rate reduced by a factor of 10 and increasing the energy loss weight to 1000.

\subsection*{Data selection with DEAL}
We employed the data efficient active learning (DEAL)~\cite{perego_data_2024} scheme to select a minimal and diverse subset of configurations for MC-PDFT labelling, used both to expand the training set of the machine learning potential and to update the WASP checkpoint library. DEAL identifies structures that are both uncertain under the current ML model and structurally non-redundant, maximizing the impact of each labeled configuration.

In the first stage, configurations are pre-filtered using a query-by-committee approach that estimates model uncertainty based on the disagreement among an ensemble of neural networks trained on the same dataset. This step ensures that only the most poorly described and informative regions of configuration space are passed on for further analysis.
In the second stage, a sparse Gaussian process (GP) is trained on the filtered configurations, using the ACE~\cite{Drautz2019AtomicPotentials} descriptors. The GP’s predictive uncertainty is used as a proxy for dissimilarity: configurations whose uncertainty is higher than a threshold are iteratively added to the GP, which become the candidates for the incremental expansion of the dataset. 

Local atomic environments are encoded using ACE-B2 descriptors calculated with the FLARE~\cite{Vandermause2020On-the-flyEvents} GP, with a basis expansion defined by $N_{\mathrm{max}} = 8$, $l_{\mathrm{max}} = 3$, and a radial cutoff of \qty{5.5}{\angstrom}, combined with a cosine switching function. Similarity is quantified using a squared normalized dot product kernel. During the initial selection from AIMD trajectories, the kernel update threshold is set to 0.004 to handle the high correlation of the dataset. For subsequent active learning rounds, the threshold is increased to 0.01–0.02 to maintain consistent selection density across rounds.

\subsection*{Incremental update of the dataset}
To maintain the consistency and accuracy of active-space assignments throughout the learning process, MC-PDFT labelling and checkpoint updates are carried out in a stepwise manner. During each round of active learning, candidate configurations produced by DEAL are first ranked based on their structural similarity to existing entries in the WASP library, quantified via the minimum RMSD. Those closest to the current library are prioritized for labelling, as they are most likely to yield reliable orbital guesses through interpolation.

To further safeguard against errors arising from extrapolation, we assess the agreement between the MLP-predicted energies and forces and their MC-PDFT counterparts. Only configurations where the energy discrepancy is less than \qty{0.08}{\eV} and the force deviation remains below \qty{0.3}{\eV\per\angstrom} are retained. This selective approach ensures that new data points are added only in regions where the model is already locally accurate and the active-space interpolation remains valid. The retained configurations are then used both to expand the checkpoint library and to retrain the machine learning potential, progressively refining its coverage of relevant configurational space while minimizing the risk of instability due to poor orbital initialization or out-of-distribution predictions. fig.~S6 shows the increase in accepted configurations over successive iterations after the dual-filter strategy supporting the robustness of the protocol.

\subsection*{Molecular dynamics and enhanced sampling}

Molecular dynamics simulations are performed using the large-scale atomic/molecular massively parallel simulator (LAMMPS) \cite{Thompson2022LAMMPSScales}, interfaced with MACE (version 0.3.7) \cite{Batatia2022MACE:Fields} and PLUMED (version 2.9) \cite{Tribello2014PLUMEDBird}. Canonical sampling is ensured via the velocity-rescaling thermostat \cite{Bussi2007CanonicalRescaling} with a coupling constant of \qty{100}{\fs}, maintaining the system at \qty{300}{\kelvin}. The integration timestep is set to \qty{0.2}{\fs}.

Enhanced sampling techniques are employed to explore reactive events efficiently. In particular, we use OPES~\cite{invernizzi_rethinking_2020}, in the flooding variant \cite{ray2022rare}. This method enhances fluctuations along a chosen collective variable (CV) that is a function of atomic coordinates, designed to represent the reaction progress.

To enable the \textbf{R}$\rightarrow$\textbf{P} transition, we use the CV defined by the difference between the coordination numbers of the two carbon atoms ($C^1$ and $C^2$) with hydrogen atoms:
\begin{equation}
    C^{\mathrm{diff}} = C^{1} - C^{2}
\end{equation}
with
\begin{equation}
    C^{i} = \sum_{j \in \ce{H}} \frac{1 -  s_{ij}^n}{1 - s_{ij}^m}
\end{equation}
\begin{equation}
    s_{ij}=\frac{r_{ij} - d_0}{r_0}
\end{equation}
where $r_{ij}$ is the distance between the carbon atom $\ce{C}_i$ and the $j$-th hydrogen atom. The parameters are set to $r_0 = $ \qty{1}{\angstrom}, $d_0 = $ \qty{0.4}{\angstrom}, $n = 6$, and $m = 12$. The bias potential is updated every 100 steps, with a maximum barrier height of \qty{42}{\kilo\joule\per\mole}. The bias deposition is limited to the region where $C^{\mathrm{diff}} > 3$ to prevent the bias from influencing the transition state and product regions.

Furthermore, to restrict the system to the reaction between $\mathbf{R}$ and $\mathbf{P}$, we apply harmonic constraints on two degrees of freedom: 1) the maximum \ce{Ti-C} distance, defined differentially as: 
\begin{equation}
    r_{\mathrm{max}} = \beta \log \sum_i \exp{\frac{r_{\ce{TiC}_i}}{\beta} }
\end{equation}
with $\beta = 0.1$, when $r_{\mathrm{max}} >$ \qty{2.75}{\angstrom}, to prevent methane desorption; 2) the angle $\theta$ between the two carbon atoms and the \ce{Ti} atom, when $\theta < $ \qty{80}{\degree}, to avoid the formation of a \ce{C-C} complex and the proceed of the reaction.

\section*{Supporting Information}
Active space used, WASP parametrization details, active learning exploration statistics, additional results from MLP-based MD simulations, optimized coordinates

\section*{Acknowledgement}
A.S., A.L.F and L.G. acknowledge support from the Catalyst Design for Decarbonization Center an Energy Frontier Research Center funded by the U.S. Department of Energy, Office of Science, Basic Energy Sciences under Award No. DE-SC0023383. M.R.H acknowledges support by the National Science Foundation Graduate Research Fellowship under Grant No. 2140001. L.B. and M.P. acknowledge funding from the European Union - NextGenerationEU initiative and the Italian National Recovery and Resilience Plan (PNRR) from the Ministry of University and Research (MUR), under Project PE0000013 CUP J53C22003010006 ``Future Artificial Intelligence Research (FAIR)''. S.P. and M.P. acknowledge funding from the Federal Ministry of Education and Research, Germany (Bundesministerium für Bildung und Forschung, BMBF, Hydrogen flagship project: TransHyDE Forschungsverbund AmmoRef, FKZ 03HY203A). We acknowledge the University of Chicago’s Research Computing Center and the CINECA award under the ISCRA initiative (project HP10B9XORR) for providing computational resources that support this work. Any opinion, findings, and conclusions or recommendations expressed in this material are those of the author(s) and do not necessarily reflect the views of the National Science Foundation.

\section*{Conflict of Interest Disclosure}

A.L.F.\ is a co-founder and consultant of Evozyne, Inc.\ and a co-author of US Patent Applications 16/887,710 and 17/642,582, US Provisional Patent Applications 62/853,919, 62/900,420, 63/314,898, 63/479,378, and 63/521,617, and International Patent Applications PCT/US2020/035206 and PCT/US2020/050466.

\bibliography{ref}

\begin{thebibliography}{76}%
\makeatletter
\providecommand \@ifxundefined [1]{%
 \@ifx{#1\undefined}
}%
\providecommand \@ifnum [1]{%
 \ifnum #1\expandafter \@firstoftwo
 \else \expandafter \@secondoftwo
 \fi
}%
\providecommand \@ifx [1]{%
 \ifx #1\expandafter \@firstoftwo
 \else \expandafter \@secondoftwo
 \fi
}%
\providecommand \natexlab [1]{#1}%
\providecommand \enquote  [1]{``#1''}%
\providecommand \bibnamefont  [1]{#1}%
\providecommand \bibfnamefont [1]{#1}%
\providecommand \citenamefont [1]{#1}%
\providecommand \href@noop [0]{\@secondoftwo}%
\providecommand \href [0]{\begingroup \@sanitize@url \@href}%
\providecommand \@href[1]{\@@startlink{#1}\@@href}%
\providecommand \@@href[1]{\endgroup#1\@@endlink}%
\providecommand \@sanitize@url [0]{\catcode `\\12\catcode `\$12\catcode `\&12\catcode `\#12\catcode `\^12\catcode `\_12\catcode `\%12\relax}%
\providecommand \@@startlink[1]{}%
\providecommand \@@endlink[0]{}%
\providecommand \url  [0]{\begingroup\@sanitize@url \@url }%
\providecommand \@url [1]{\endgroup\@href {#1}{\urlprefix }}%
\providecommand \urlprefix  [0]{URL }%
\providecommand \Eprint [0]{\href }%
\providecommand \doibase [0]{https://doi.org/}%
\providecommand \selectlanguage [0]{\@gobble}%
\providecommand \bibinfo  [0]{\@secondoftwo}%
\providecommand \bibfield  [0]{\@secondoftwo}%
\providecommand \translation [1]{[#1]}%
\providecommand \BibitemOpen [0]{}%
\providecommand \bibitemStop [0]{}%
\providecommand \bibitemNoStop [0]{.\EOS\space}%
\providecommand \EOS [0]{\spacefactor3000\relax}%
\providecommand \BibitemShut  [1]{\csname bibitem#1\endcsname}%
\let\auto@bib@innerbib\@empty
%</preamble>
\bibitem [{\citenamefont {Deringer}\ \emph {et~al.}(2019)\citenamefont {Deringer}, \citenamefont {Caro},\ and\ \citenamefont {Csányi}}]{deringer_machine_2019}%
  \BibitemOpen
  \bibfield  {author} {\bibinfo {author} {\bibfnamefont {V.~L.}\ \bibnamefont {Deringer}}, \bibinfo {author} {\bibfnamefont {M.~A.}\ \bibnamefont {Caro}},\ and\ \bibinfo {author} {\bibfnamefont {G.}~\bibnamefont {Csányi}},\ }\href {https://doi.org/10.1002/adma.201902765} {\bibfield  {journal} {\bibinfo  {journal} {Adv. Mater.}\ }\textbf {\bibinfo {volume} {31}},\ \bibinfo {pages} {1902765} (\bibinfo {year} {2019})}\BibitemShut {NoStop}%
\bibitem [{\citenamefont {Schran}\ \emph {et~al.}(2021)\citenamefont {Schran}, \citenamefont {Thiemann}, \citenamefont {Rowe}, \citenamefont {Müller}, \citenamefont {Marsalek},\ and\ \citenamefont {Michaelides}}]{schran_machine_2021}%
  \BibitemOpen
  \bibfield  {author} {\bibinfo {author} {\bibfnamefont {C.}~\bibnamefont {Schran}}, \bibinfo {author} {\bibfnamefont {F.~L.}\ \bibnamefont {Thiemann}}, \bibinfo {author} {\bibfnamefont {P.}~\bibnamefont {Rowe}}, \bibinfo {author} {\bibfnamefont {E.~A.}\ \bibnamefont {Müller}}, \bibinfo {author} {\bibfnamefont {O.}~\bibnamefont {Marsalek}},\ and\ \bibinfo {author} {\bibfnamefont {A.}~\bibnamefont {Michaelides}},\ }\href {https://doi.org/10.1073/pnas.2110077118} {\bibfield  {journal} {\bibinfo  {journal} {Proc. Natl. Acad. Sci. U. S. A.}\ }\textbf {\bibinfo {volume} {118}},\ \bibinfo {pages} {e2110077118} (\bibinfo {year} {2021})}\BibitemShut {NoStop}%
\bibitem [{\citenamefont {Unke}\ \emph {et~al.}(2021)\citenamefont {Unke}, \citenamefont {Chmiela}, \citenamefont {Sauceda}, \citenamefont {Gastegger}, \citenamefont {Poltavsky}, \citenamefont {Schütt}, \citenamefont {Tkatchenko},\ and\ \citenamefont {Müller}}]{unke_machine_2021}%
  \BibitemOpen
  \bibfield  {author} {\bibinfo {author} {\bibfnamefont {O.~T.}\ \bibnamefont {Unke}}, \bibinfo {author} {\bibfnamefont {S.}~\bibnamefont {Chmiela}}, \bibinfo {author} {\bibfnamefont {H.~E.}\ \bibnamefont {Sauceda}}, \bibinfo {author} {\bibfnamefont {M.}~\bibnamefont {Gastegger}}, \bibinfo {author} {\bibfnamefont {I.}~\bibnamefont {Poltavsky}}, \bibinfo {author} {\bibfnamefont {K.~T.}\ \bibnamefont {Schütt}}, \bibinfo {author} {\bibfnamefont {A.}~\bibnamefont {Tkatchenko}},\ and\ \bibinfo {author} {\bibfnamefont {K.-R.}\ \bibnamefont {Müller}},\ }\href {https://doi.org/10.1021/acs.chemrev.0c01111} {\bibfield  {journal} {\bibinfo  {journal} {Chem. Rev.}\ }\textbf {\bibinfo {volume} {121}},\ \bibinfo {pages} {10142} (\bibinfo {year} {2021})}\BibitemShut {NoStop}%
\bibitem [{\citenamefont {Bonati}\ \emph {et~al.}(2023)\citenamefont {Bonati}, \citenamefont {Polino}, \citenamefont {Pizzolitto}, \citenamefont {Biasi}, \citenamefont {Eckert}, \citenamefont {Reitmeier}, \citenamefont {Schlögl},\ and\ \citenamefont {Parrinello}}]{bonati_role_2023}%
  \BibitemOpen
  \bibfield  {author} {\bibinfo {author} {\bibfnamefont {L.}~\bibnamefont {Bonati}}, \bibinfo {author} {\bibfnamefont {D.}~\bibnamefont {Polino}}, \bibinfo {author} {\bibfnamefont {C.}~\bibnamefont {Pizzolitto}}, \bibinfo {author} {\bibfnamefont {P.}~\bibnamefont {Biasi}}, \bibinfo {author} {\bibfnamefont {R.}~\bibnamefont {Eckert}}, \bibinfo {author} {\bibfnamefont {S.}~\bibnamefont {Reitmeier}}, \bibinfo {author} {\bibfnamefont {R.}~\bibnamefont {Schlögl}},\ and\ \bibinfo {author} {\bibfnamefont {M.}~\bibnamefont {Parrinello}},\ }\href {https://doi.org/10.1073/pnas.2313023120} {\bibfield  {journal} {\bibinfo  {journal} {Proc. Natl. Acad. Sci. U. S. A.}\ }\textbf {\bibinfo {volume} {120}},\ \bibinfo {pages} {e2313023120} (\bibinfo {year} {2023})}\BibitemShut {NoStop}%
\bibitem [{\citenamefont {Yang}\ \emph {et~al.}(2023)\citenamefont {Yang}, \citenamefont {Raucci},\ and\ \citenamefont {Parrinello}}]{yang2023reactant}%
  \BibitemOpen
  \bibfield  {author} {\bibinfo {author} {\bibfnamefont {M.}~\bibnamefont {Yang}}, \bibinfo {author} {\bibfnamefont {U.}~\bibnamefont {Raucci}},\ and\ \bibinfo {author} {\bibfnamefont {M.}~\bibnamefont {Parrinello}},\ }\href {https://doi.org/10.1038/s41929-023-01006-2} {\bibfield  {journal} {\bibinfo  {journal} {Nat. Catal.}\ }\textbf {\bibinfo {volume} {6}},\ \bibinfo {pages} {829} (\bibinfo {year} {2023})}\BibitemShut {NoStop}%
\bibitem [{\citenamefont {Perego}\ \emph {et~al.}(2024)\citenamefont {Perego}, \citenamefont {Bonati}, \citenamefont {Tripathi},\ and\ \citenamefont {Parrinello}}]{perego2024dynamics}%
  \BibitemOpen
  \bibfield  {author} {\bibinfo {author} {\bibfnamefont {S.}~\bibnamefont {Perego}}, \bibinfo {author} {\bibfnamefont {L.}~\bibnamefont {Bonati}}, \bibinfo {author} {\bibfnamefont {S.}~\bibnamefont {Tripathi}},\ and\ \bibinfo {author} {\bibfnamefont {M.}~\bibnamefont {Parrinello}},\ }\href {https://doi.org/10.1021/acscatal.4c01920} {\bibfield  {journal} {\bibinfo  {journal} {ACS Catal.}\ }\textbf {\bibinfo {volume} {14}},\ \bibinfo {pages} {14652} (\bibinfo {year} {2024})}\BibitemShut {NoStop}%
\bibitem [{\citenamefont {Omranpour}\ \emph {et~al.}(2025)\citenamefont {Omranpour}, \citenamefont {Elsner}, \citenamefont {Lausch},\ and\ \citenamefont {Behler}}]{omranpour_machine_2025}%
  \BibitemOpen
  \bibfield  {author} {\bibinfo {author} {\bibfnamefont {A.}~\bibnamefont {Omranpour}}, \bibinfo {author} {\bibfnamefont {J.}~\bibnamefont {Elsner}}, \bibinfo {author} {\bibfnamefont {K.~N.}\ \bibnamefont {Lausch}},\ and\ \bibinfo {author} {\bibfnamefont {J.}~\bibnamefont {Behler}},\ }\href {https://doi.org/10.1021/acscatal.4c06717} {\bibfield  {journal} {\bibinfo  {journal} {ACS Catal.}\ }\textbf {\bibinfo {volume} {15}},\ \bibinfo {pages} {1616} (\bibinfo {year} {2025})}\BibitemShut {NoStop}%
\bibitem [{\citenamefont {Bonati}\ and\ \citenamefont {Parrinello}(2018)}]{bonati2018silicon}%
  \BibitemOpen
  \bibfield  {author} {\bibinfo {author} {\bibfnamefont {L.}~\bibnamefont {Bonati}}\ and\ \bibinfo {author} {\bibfnamefont {M.}~\bibnamefont {Parrinello}},\ }\href {https://doi.org/10.1103/PhysRevLett.121.265701} {\bibfield  {journal} {\bibinfo  {journal} {Phys. Rev. Lett.}\ }\textbf {\bibinfo {volume} {121}},\ \bibinfo {pages} {265701} (\bibinfo {year} {2018})}\BibitemShut {NoStop}%
\bibitem [{\citenamefont {Jinnouchi}\ \emph {et~al.}(2019)\citenamefont {Jinnouchi}, \citenamefont {Lahnsteiner}, \citenamefont {Karsai}, \citenamefont {Kresse},\ and\ \citenamefont {Bokdam}}]{jinnouchi_phase_2019}%
  \BibitemOpen
  \bibfield  {author} {\bibinfo {author} {\bibfnamefont {R.}~\bibnamefont {Jinnouchi}}, \bibinfo {author} {\bibfnamefont {J.}~\bibnamefont {Lahnsteiner}}, \bibinfo {author} {\bibfnamefont {F.}~\bibnamefont {Karsai}}, \bibinfo {author} {\bibfnamefont {G.}~\bibnamefont {Kresse}},\ and\ \bibinfo {author} {\bibfnamefont {M.}~\bibnamefont {Bokdam}},\ }\href {https://doi.org/10.1103/PhysRevLett.122.225701} {\bibfield  {journal} {\bibinfo  {journal} {Phys. Rev. Lett.}\ }\textbf {\bibinfo {volume} {122}},\ \bibinfo {pages} {225701} (\bibinfo {year} {2019})}\BibitemShut {NoStop}%
\bibitem [{\citenamefont {Deringer}\ \emph {et~al.}(2021)\citenamefont {Deringer}, \citenamefont {Bernstein}, \citenamefont {Csányi}, \citenamefont {Ben~Mahmoud}, \citenamefont {Ceriotti}, \citenamefont {Wilson}, \citenamefont {Drabold},\ and\ \citenamefont {Elliott}}]{deringer_origins_2021}%
  \BibitemOpen
  \bibfield  {author} {\bibinfo {author} {\bibfnamefont {V.~L.}\ \bibnamefont {Deringer}}, \bibinfo {author} {\bibfnamefont {N.}~\bibnamefont {Bernstein}}, \bibinfo {author} {\bibfnamefont {G.}~\bibnamefont {Csányi}}, \bibinfo {author} {\bibfnamefont {C.}~\bibnamefont {Ben~Mahmoud}}, \bibinfo {author} {\bibfnamefont {M.}~\bibnamefont {Ceriotti}}, \bibinfo {author} {\bibfnamefont {M.}~\bibnamefont {Wilson}}, \bibinfo {author} {\bibfnamefont {D.~A.}\ \bibnamefont {Drabold}},\ and\ \bibinfo {author} {\bibfnamefont {S.~R.}\ \bibnamefont {Elliott}},\ }\href {https://doi.org/10.1038/s41586-020-03072-z} {\bibfield  {journal} {\bibinfo  {journal} {Nature}\ }\textbf {\bibinfo {volume} {589}},\ \bibinfo {pages} {59} (\bibinfo {year} {2021})}\BibitemShut {NoStop}%
\bibitem [{\citenamefont {Abou El~Kheir}\ \emph {et~al.}(2024)\citenamefont {Abou El~Kheir}, \citenamefont {Bonati}, \citenamefont {Parrinello},\ and\ \citenamefont {Bernasconi}}]{abou2024unraveling}%
  \BibitemOpen
  \bibfield  {author} {\bibinfo {author} {\bibfnamefont {O.}~\bibnamefont {Abou El~Kheir}}, \bibinfo {author} {\bibfnamefont {L.}~\bibnamefont {Bonati}}, \bibinfo {author} {\bibfnamefont {M.}~\bibnamefont {Parrinello}},\ and\ \bibinfo {author} {\bibfnamefont {M.}~\bibnamefont {Bernasconi}},\ }\href {https://doi.org/10.1038/s41524-024-01217-6} {\bibfield  {journal} {\bibinfo  {journal} {Npj Comput. Mater.}\ }\textbf {\bibinfo {volume} {10}},\ \bibinfo {pages} {1} (\bibinfo {year} {2024})}\BibitemShut {NoStop}%
\bibitem [{\citenamefont {Dral}\ and\ \citenamefont {Barbatti}(2021)}]{dral2021molecular}%
  \BibitemOpen
  \bibfield  {author} {\bibinfo {author} {\bibfnamefont {P.~O.}\ \bibnamefont {Dral}}\ and\ \bibinfo {author} {\bibfnamefont {M.}~\bibnamefont {Barbatti}},\ }\href {https://doi.org/10.1038/s41570-021-00278-1} {\bibfield  {journal} {\bibinfo  {journal} {Nat. Rev. Chem.}\ }\textbf {\bibinfo {volume} {5}},\ \bibinfo {pages} {388} (\bibinfo {year} {2021})}\BibitemShut {NoStop}%
\bibitem [{\citenamefont {Raucci}(2025)}]{raucci2025capturing}%
  \BibitemOpen
  \bibfield  {author} {\bibinfo {author} {\bibfnamefont {U.}~\bibnamefont {Raucci}},\ }\href {https://doi.org/10.1021/acs.jpclett.5c00688} {\bibfield  {journal} {\bibinfo  {journal} {J. Phys. Chem. Lett.}\ ,\ \bibinfo {pages} {4900}} (\bibinfo {year} {2025})}\BibitemShut {NoStop}%
\bibitem [{\citenamefont {Westermayr}\ and\ \citenamefont {Marquetand}(2021)}]{westermayr2020machine}%
  \BibitemOpen
  \bibfield  {author} {\bibinfo {author} {\bibfnamefont {J.}~\bibnamefont {Westermayr}}\ and\ \bibinfo {author} {\bibfnamefont {P.}~\bibnamefont {Marquetand}},\ }\href {https://doi.org/10.1021/acs.chemrev.0c00749} {\bibfield  {journal} {\bibinfo  {journal} {Chem. Rev.}\ }\textbf {\bibinfo {volume} {121}},\ \bibinfo {pages} {9873} (\bibinfo {year} {2021})}\BibitemShut {NoStop}%
\bibitem [{\citenamefont {Kulichenko}\ \emph {et~al.}(2024)\citenamefont {Kulichenko}, \citenamefont {Nebgen}, \citenamefont {Lubbers}, \citenamefont {Smith}, \citenamefont {Barros}, \citenamefont {Allen}, \citenamefont {Habib}, \citenamefont {Shinkle}, \citenamefont {Fedik}, \citenamefont {Li}, \citenamefont {Messerly},\ and\ \citenamefont {Tretiak}}]{kulichenko_data_2024}%
  \BibitemOpen
  \bibfield  {author} {\bibinfo {author} {\bibfnamefont {M.}~\bibnamefont {Kulichenko}}, \bibinfo {author} {\bibfnamefont {B.}~\bibnamefont {Nebgen}}, \bibinfo {author} {\bibfnamefont {N.}~\bibnamefont {Lubbers}}, \bibinfo {author} {\bibfnamefont {J.~S.}\ \bibnamefont {Smith}}, \bibinfo {author} {\bibfnamefont {K.}~\bibnamefont {Barros}}, \bibinfo {author} {\bibfnamefont {A.~E.~A.}\ \bibnamefont {Allen}}, \bibinfo {author} {\bibfnamefont {A.}~\bibnamefont {Habib}}, \bibinfo {author} {\bibfnamefont {E.}~\bibnamefont {Shinkle}}, \bibinfo {author} {\bibfnamefont {N.}~\bibnamefont {Fedik}}, \bibinfo {author} {\bibfnamefont {Y.~W.}\ \bibnamefont {Li}}, \bibinfo {author} {\bibfnamefont {R.~A.}\ \bibnamefont {Messerly}},\ and\ \bibinfo {author} {\bibfnamefont {S.}~\bibnamefont {Tretiak}},\ }\href {https://doi.org/10.1021/acs.chemrev.4c00572} {\bibfield  {journal} {\bibinfo  {journal} {Chem. Rev.}\ }\textbf {\bibinfo {volume} {124}},\ \bibinfo {pages} {13681} (\bibinfo {year} {2024})}\BibitemShut {NoStop}%
\bibitem [{\citenamefont {Christensen}\ \emph {et~al.}(2021)\citenamefont {Christensen}, \citenamefont {Sirumalla}, \citenamefont {Qiao}, \citenamefont {O’Connor}, \citenamefont {Smith}, \citenamefont {Ding}, \citenamefont {Bygrave}, \citenamefont {Anandkumar}, \citenamefont {Welborn}, \citenamefont {Manby},\ and\ \citenamefont {Miller}}]{christensen_orbnet_2021}%
  \BibitemOpen
  \bibfield  {author} {\bibinfo {author} {\bibfnamefont {A.~S.}\ \bibnamefont {Christensen}}, \bibinfo {author} {\bibfnamefont {S.~K.}\ \bibnamefont {Sirumalla}}, \bibinfo {author} {\bibfnamefont {Z.}~\bibnamefont {Qiao}}, \bibinfo {author} {\bibfnamefont {M.~B.}\ \bibnamefont {O’Connor}}, \bibinfo {author} {\bibfnamefont {D.~G.~A.}\ \bibnamefont {Smith}}, \bibinfo {author} {\bibfnamefont {F.}~\bibnamefont {Ding}}, \bibinfo {author} {\bibfnamefont {P.~J.}\ \bibnamefont {Bygrave}}, \bibinfo {author} {\bibfnamefont {A.}~\bibnamefont {Anandkumar}}, \bibinfo {author} {\bibfnamefont {M.}~\bibnamefont {Welborn}}, \bibinfo {author} {\bibfnamefont {F.~R.}\ \bibnamefont {Manby}},\ and\ \bibinfo {author} {\bibfnamefont {T.~F.}\ \bibnamefont {Miller}, \bibfnamefont {III}},\ }\href {https://doi.org/10.1063/5.0061990} {\bibfield  {journal} {\bibinfo  {journal} {J. Chem. Phys.}\ }\textbf {\bibinfo {volume} {155}},\ \bibinfo {pages} {204103} (\bibinfo {year} {2021})}\BibitemShut {NoStop}%
\bibitem [{\citenamefont {Wang}\ \emph {et~al.}(2024)\citenamefont {Wang}, \citenamefont {He}, \citenamefont {Li}, \citenamefont {Li}, \citenamefont {Bi}, \citenamefont {Wang}, \citenamefont {Cheng}, \citenamefont {Shen}, \citenamefont {Meng}, \citenamefont {Zhang}, \citenamefont {Liu}, \citenamefont {Wang}, \citenamefont {Li}, \citenamefont {Shao},\ and\ \citenamefont {Liu}}]{wang_ab_2024}%
  \BibitemOpen
  \bibfield  {author} {\bibinfo {author} {\bibfnamefont {T.}~\bibnamefont {Wang}}, \bibinfo {author} {\bibfnamefont {X.}~\bibnamefont {He}}, \bibinfo {author} {\bibfnamefont {M.}~\bibnamefont {Li}}, \bibinfo {author} {\bibfnamefont {Y.}~\bibnamefont {Li}}, \bibinfo {author} {\bibfnamefont {R.}~\bibnamefont {Bi}}, \bibinfo {author} {\bibfnamefont {Y.}~\bibnamefont {Wang}}, \bibinfo {author} {\bibfnamefont {C.}~\bibnamefont {Cheng}}, \bibinfo {author} {\bibfnamefont {X.}~\bibnamefont {Shen}}, \bibinfo {author} {\bibfnamefont {J.}~\bibnamefont {Meng}}, \bibinfo {author} {\bibfnamefont {H.}~\bibnamefont {Zhang}}, \bibinfo {author} {\bibfnamefont {H.}~\bibnamefont {Liu}}, \bibinfo {author} {\bibfnamefont {Z.}~\bibnamefont {Wang}}, \bibinfo {author} {\bibfnamefont {S.}~\bibnamefont {Li}}, \bibinfo {author} {\bibfnamefont {B.}~\bibnamefont {Shao}},\ and\ \bibinfo {author} {\bibfnamefont {T.-Y.}\ \bibnamefont {Liu}},\ }\href {https://doi.org/10.1038/s41586-024-08127-z} {\bibfield  {journal} {\bibinfo  {journal}
  {Nature}\ }\textbf {\bibinfo {volume} {635}},\ \bibinfo {pages} {1019} (\bibinfo {year} {2024})}\BibitemShut {NoStop}%
\bibitem [{\citenamefont {Roos}\ \emph {et~al.}(1980)\citenamefont {Roos}, \citenamefont {Taylor},\ and\ \citenamefont {Sigbahn}}]{roos_complete_1980}%
  \BibitemOpen
  \bibfield  {author} {\bibinfo {author} {\bibfnamefont {B.~O.}\ \bibnamefont {Roos}}, \bibinfo {author} {\bibfnamefont {P.~R.}\ \bibnamefont {Taylor}},\ and\ \bibinfo {author} {\bibfnamefont {P.~E.~M.}\ \bibnamefont {Sigbahn}},\ }\href {https://doi.org/10.1016/0301-0104(80)80045-0} {\bibfield  {journal} {\bibinfo  {journal} {Chem. Phys.}\ }\textbf {\bibinfo {volume} {48}},\ \bibinfo {pages} {157} (\bibinfo {year} {1980})}\BibitemShut {NoStop}%
\bibitem [{\citenamefont {Roos}(1987)}]{roos_complete_1987}%
  \BibitemOpen
  \bibfield  {author} {\bibinfo {author} {\bibfnamefont {B.~O.}\ \bibnamefont {Roos}},\ }in\ \href {https://onlinelibrary.wiley.com/doi/abs/10.1002/9780470142943.ch7} {\emph {\bibinfo {booktitle} {Advances in {Chemical} {Physics}}}}\ (\bibinfo  {publisher} {John Wiley \& Sons, Ltd},\ \bibinfo {year} {1987})\ pp.\ \bibinfo {pages} {399--445}\BibitemShut {NoStop}%
\bibitem [{\citenamefont {Li~Manni}\ \emph {et~al.}(2014)\citenamefont {Li~Manni}, \citenamefont {Carlson}, \citenamefont {Luo}, \citenamefont {Ma}, \citenamefont {Olsen}, \citenamefont {Truhlar},\ and\ \citenamefont {Gagliardi}}]{li_manni_multiconfiguration_2014}%
  \BibitemOpen
  \bibfield  {author} {\bibinfo {author} {\bibfnamefont {G.}~\bibnamefont {Li~Manni}}, \bibinfo {author} {\bibfnamefont {R.~K.}\ \bibnamefont {Carlson}}, \bibinfo {author} {\bibfnamefont {S.}~\bibnamefont {Luo}}, \bibinfo {author} {\bibfnamefont {D.}~\bibnamefont {Ma}}, \bibinfo {author} {\bibfnamefont {J.}~\bibnamefont {Olsen}}, \bibinfo {author} {\bibfnamefont {D.~G.}\ \bibnamefont {Truhlar}},\ and\ \bibinfo {author} {\bibfnamefont {L.}~\bibnamefont {Gagliardi}},\ }\href {https://doi.org/10.1021/ct500483t} {\bibfield  {journal} {\bibinfo  {journal} {J. Chem. Theory Comput.}\ }\textbf {\bibinfo {volume} {10}},\ \bibinfo {pages} {3669} (\bibinfo {year} {2014})}\BibitemShut {NoStop}%
\bibitem [{\citenamefont {Sharma}\ \emph {et~al.}(2021)\citenamefont {Sharma}, \citenamefont {Bao}, \citenamefont {Truhlar},\ and\ \citenamefont {Gagliardi}}]{sharma_multiconfiguration_2021}%
  \BibitemOpen
  \bibfield  {author} {\bibinfo {author} {\bibfnamefont {P.}~\bibnamefont {Sharma}}, \bibinfo {author} {\bibfnamefont {J.~J.}\ \bibnamefont {Bao}}, \bibinfo {author} {\bibfnamefont {D.~G.}\ \bibnamefont {Truhlar}},\ and\ \bibinfo {author} {\bibfnamefont {L.}~\bibnamefont {Gagliardi}},\ }\href {https://doi.org/10.1146/annurev-physchem-090419-043839} {\bibfield  {journal} {\bibinfo  {journal} {Annu. Rev. Phys. Chem.}\ }\textbf {\bibinfo {volume} {72}},\ \bibinfo {pages} {541} (\bibinfo {year} {2021})}\BibitemShut {NoStop}%
\bibitem [{\citenamefont {Zhou}\ \emph {et~al.}(2022)\citenamefont {Zhou}, \citenamefont {Hermes}, \citenamefont {Wu}, \citenamefont {Bao}, \citenamefont {Pandharkar}, \citenamefont {King}, \citenamefont {Zhang}, \citenamefont {Scott}, \citenamefont {Lykhin}, \citenamefont {Gagliardi},\ and\ \citenamefont {Truhlar}}]{zhou_electronic_2022}%
  \BibitemOpen
  \bibfield  {author} {\bibinfo {author} {\bibfnamefont {C.}~\bibnamefont {Zhou}}, \bibinfo {author} {\bibfnamefont {M.~R.}\ \bibnamefont {Hermes}}, \bibinfo {author} {\bibfnamefont {D.}~\bibnamefont {Wu}}, \bibinfo {author} {\bibfnamefont {J.~J.}\ \bibnamefont {Bao}}, \bibinfo {author} {\bibfnamefont {R.}~\bibnamefont {Pandharkar}}, \bibinfo {author} {\bibfnamefont {D.~S.}\ \bibnamefont {King}}, \bibinfo {author} {\bibfnamefont {D.}~\bibnamefont {Zhang}}, \bibinfo {author} {\bibfnamefont {T.~R.}\ \bibnamefont {Scott}}, \bibinfo {author} {\bibfnamefont {A.~O.}\ \bibnamefont {Lykhin}}, \bibinfo {author} {\bibfnamefont {L.}~\bibnamefont {Gagliardi}},\ and\ \bibinfo {author} {\bibfnamefont {D.~G.}\ \bibnamefont {Truhlar}},\ }\href {https://doi.org/10.1039/D2SC01022D} {\bibfield  {journal} {\bibinfo  {journal} {Chemical Science}\ }\textbf {\bibinfo {volume} {13}},\ \bibinfo {pages} {7685} (\bibinfo {year} {2022})}\BibitemShut {NoStop}%
\bibitem [{\citenamefont {Angeli}\ \emph {et~al.}(2001)\citenamefont {Angeli}, \citenamefont {Cimiraglia}, \citenamefont {Evangelisti}, \citenamefont {Leininger},\ and\ \citenamefont {Malrieu}}]{angeli_introduction_2001}%
  \BibitemOpen
  \bibfield  {author} {\bibinfo {author} {\bibfnamefont {C.}~\bibnamefont {Angeli}}, \bibinfo {author} {\bibfnamefont {R.}~\bibnamefont {Cimiraglia}}, \bibinfo {author} {\bibfnamefont {S.}~\bibnamefont {Evangelisti}}, \bibinfo {author} {\bibfnamefont {T.}~\bibnamefont {Leininger}},\ and\ \bibinfo {author} {\bibfnamefont {J.-P.}\ \bibnamefont {Malrieu}},\ }\href {https://doi.org/10.1063/1.1361246} {\bibfield  {journal} {\bibinfo  {journal} {J. Chem. Phys.}\ }\textbf {\bibinfo {volume} {114}},\ \bibinfo {pages} {10252} (\bibinfo {year} {2001})}\BibitemShut {NoStop}%
\bibitem [{\citenamefont {Andersson}\ \emph {et~al.}(1990)\citenamefont {Andersson}, \citenamefont {Malmqvist}, \citenamefont {Roos}, \citenamefont {Sadlej},\ and\ \citenamefont {Wolinski}}]{andersson_second-order_1990}%
  \BibitemOpen
  \bibfield  {author} {\bibinfo {author} {\bibfnamefont {K.}~\bibnamefont {Andersson}}, \bibinfo {author} {\bibfnamefont {P.~A.}\ \bibnamefont {Malmqvist}}, \bibinfo {author} {\bibfnamefont {B.~O.}\ \bibnamefont {Roos}}, \bibinfo {author} {\bibfnamefont {A.~J.}\ \bibnamefont {Sadlej}},\ and\ \bibinfo {author} {\bibfnamefont {K.}~\bibnamefont {Wolinski}},\ }\href {https://doi.org/10.1021/j100377a012} {\bibfield  {journal} {\bibinfo  {journal} {J. Phys. Chem.}\ }\textbf {\bibinfo {volume} {94}},\ \bibinfo {pages} {5483} (\bibinfo {year} {1990})}\BibitemShut {NoStop}%
\bibitem [{\citenamefont {Hoyer}\ \emph {et~al.}(2016)\citenamefont {Hoyer}, \citenamefont {Ghosh}, \citenamefont {Truhlar},\ and\ \citenamefont {Gagliardi}}]{hoyer_multiconfiguration_2016}%
  \BibitemOpen
  \bibfield  {author} {\bibinfo {author} {\bibfnamefont {C.~E.}\ \bibnamefont {Hoyer}}, \bibinfo {author} {\bibfnamefont {S.}~\bibnamefont {Ghosh}}, \bibinfo {author} {\bibfnamefont {D.~G.}\ \bibnamefont {Truhlar}},\ and\ \bibinfo {author} {\bibfnamefont {L.}~\bibnamefont {Gagliardi}},\ }\href {https://doi.org/10.1021/acs.jpclett.5b02773} {\bibfield  {journal} {\bibinfo  {journal} {J. Phys. Chem. Lett.}\ }\textbf {\bibinfo {volume} {7}},\ \bibinfo {pages} {586} (\bibinfo {year} {2016})}\BibitemShut {NoStop}%
\bibitem [{\citenamefont {King}\ \emph {et~al.}(2022)\citenamefont {King}, \citenamefont {Hermes}, \citenamefont {Truhlar},\ and\ \citenamefont {Gagliardi}}]{king_large-scale_2022}%
  \BibitemOpen
  \bibfield  {author} {\bibinfo {author} {\bibfnamefont {D.~S.}\ \bibnamefont {King}}, \bibinfo {author} {\bibfnamefont {M.~R.}\ \bibnamefont {Hermes}}, \bibinfo {author} {\bibfnamefont {D.~G.}\ \bibnamefont {Truhlar}},\ and\ \bibinfo {author} {\bibfnamefont {L.}~\bibnamefont {Gagliardi}},\ }\href {https://doi.org/10.1021/acs.jctc.2c00630} {\bibfield  {journal} {\bibinfo  {journal} {J. Chem. Theory Comput.}\ }\textbf {\bibinfo {volume} {18}},\ \bibinfo {pages} {6065} (\bibinfo {year} {2022})}\BibitemShut {NoStop}%
\bibitem [{\citenamefont {Carlson}\ \emph {et~al.}(2015)\citenamefont {Carlson}, \citenamefont {Li~Manni}, \citenamefont {Sonnenberger}, \citenamefont {Truhlar},\ and\ \citenamefont {Gagliardi}}]{carlson_multiconfiguration_2015}%
  \BibitemOpen
  \bibfield  {author} {\bibinfo {author} {\bibfnamefont {R.~K.}\ \bibnamefont {Carlson}}, \bibinfo {author} {\bibfnamefont {G.}~\bibnamefont {Li~Manni}}, \bibinfo {author} {\bibfnamefont {A.~L.}\ \bibnamefont {Sonnenberger}}, \bibinfo {author} {\bibfnamefont {D.~G.}\ \bibnamefont {Truhlar}},\ and\ \bibinfo {author} {\bibfnamefont {L.}~\bibnamefont {Gagliardi}},\ }\href {https://doi.org/10.1021/ct5008235} {\bibfield  {journal} {\bibinfo  {journal} {J. Chem. Theory Comput.}\ }\textbf {\bibinfo {volume} {11}},\ \bibinfo {pages} {82} (\bibinfo {year} {2015})}\BibitemShut {NoStop}%
\bibitem [{\citenamefont {Wardzala}\ \emph {et~al.}(2024)\citenamefont {Wardzala}, \citenamefont {King}, \citenamefont {Ogunfowora}, \citenamefont {Savoie},\ and\ \citenamefont {Gagliardi}}]{wardzala_organic_2024}%
  \BibitemOpen
  \bibfield  {author} {\bibinfo {author} {\bibfnamefont {J.~J.}\ \bibnamefont {Wardzala}}, \bibinfo {author} {\bibfnamefont {D.~S.}\ \bibnamefont {King}}, \bibinfo {author} {\bibfnamefont {L.}~\bibnamefont {Ogunfowora}}, \bibinfo {author} {\bibfnamefont {B.}~\bibnamefont {Savoie}},\ and\ \bibinfo {author} {\bibfnamefont {L.}~\bibnamefont {Gagliardi}},\ }\href {https://doi.org/10.1021/acscentsci.3c01559} {\bibfield  {journal} {\bibinfo  {journal} {ACS Cent. Sci.}\ }\textbf {\bibinfo {volume} {10}},\ \bibinfo {pages} {833} (\bibinfo {year} {2024})}\BibitemShut {NoStop}%
\bibitem [{\citenamefont {Mitchell}\ \emph {et~al.}(2022)\citenamefont {Mitchell}, \citenamefont {Scott}, \citenamefont {Bao},\ and\ \citenamefont {Truhlar}}]{mitchell_application_2022}%
  \BibitemOpen
  \bibfield  {author} {\bibinfo {author} {\bibfnamefont {E.~C.}\ \bibnamefont {Mitchell}}, \bibinfo {author} {\bibfnamefont {T.~R.}\ \bibnamefont {Scott}}, \bibinfo {author} {\bibfnamefont {J.~J.}\ \bibnamefont {Bao}},\ and\ \bibinfo {author} {\bibfnamefont {D.~G.}\ \bibnamefont {Truhlar}},\ }\href {https://doi.org/10.1021/acs.jpca.2c06433} {\bibfield  {journal} {\bibinfo  {journal} {J. Phys. Chem. A}\ }\textbf {\bibinfo {volume} {126}},\ \bibinfo {pages} {8834} (\bibinfo {year} {2022})}\BibitemShut {NoStop}%
\bibitem [{\citenamefont {Maity}\ \emph {et~al.}(2022)\citenamefont {Maity}, \citenamefont {Scott}, \citenamefont {Stroscio}, \citenamefont {Gagliardi},\ and\ \citenamefont {Cavallo}}]{maity_role_2022}%
  \BibitemOpen
  \bibfield  {author} {\bibinfo {author} {\bibfnamefont {B.}~\bibnamefont {Maity}}, \bibinfo {author} {\bibfnamefont {T.~R.}\ \bibnamefont {Scott}}, \bibinfo {author} {\bibfnamefont {G.~D.}\ \bibnamefont {Stroscio}}, \bibinfo {author} {\bibfnamefont {L.}~\bibnamefont {Gagliardi}},\ and\ \bibinfo {author} {\bibfnamefont {L.}~\bibnamefont {Cavallo}},\ }\href {https://doi.org/10.1021/acscatal.2c04284} {\bibfield  {journal} {\bibinfo  {journal} {ACS Catal.}\ }\textbf {\bibinfo {volume} {12}},\ \bibinfo {pages} {13215} (\bibinfo {year} {2022})}\BibitemShut {NoStop}%
\bibitem [{\citenamefont {Calio}\ \emph {et~al.}(2022)\citenamefont {Calio}, \citenamefont {Truhlar},\ and\ \citenamefont {Gagliardi}}]{calio_nonadiabatic_2022}%
  \BibitemOpen
  \bibfield  {author} {\bibinfo {author} {\bibfnamefont {P.~B.}\ \bibnamefont {Calio}}, \bibinfo {author} {\bibfnamefont {D.~G.}\ \bibnamefont {Truhlar}},\ and\ \bibinfo {author} {\bibfnamefont {L.}~\bibnamefont {Gagliardi}},\ }\href {https://doi.org/10.1021/acs.jctc.1c01048} {\bibfield  {journal} {\bibinfo  {journal} {J. Chem. Theory Comput.}\ }\textbf {\bibinfo {volume} {18}},\ \bibinfo {pages} {614} (\bibinfo {year} {2022})}\BibitemShut {NoStop}%
\bibitem [{\citenamefont {Hennefarth}\ \emph {et~al.}(2024)\citenamefont {Hennefarth}, \citenamefont {Truhlar},\ and\ \citenamefont {Gagliardi}}]{hennefarth_semiclassical_2024}%
  \BibitemOpen
  \bibfield  {author} {\bibinfo {author} {\bibfnamefont {M.~R.}\ \bibnamefont {Hennefarth}}, \bibinfo {author} {\bibfnamefont {D.~G.}\ \bibnamefont {Truhlar}},\ and\ \bibinfo {author} {\bibfnamefont {L.}~\bibnamefont {Gagliardi}},\ }\href {https://doi.org/10.1021/acs.jctc.4c01061} {\bibfield  {journal} {\bibinfo  {journal} {J. Chem. Theory Comput.}\ }\textbf {\bibinfo {volume} {20}},\ \bibinfo {pages} {8741} (\bibinfo {year} {2024})}\BibitemShut {NoStop}%
\bibitem [{\citenamefont {Scott}\ \emph {et~al.}(2021)\citenamefont {Scott}, \citenamefont {Oakley}, \citenamefont {Hermes}, \citenamefont {Sand}, \citenamefont {Lindh}, \citenamefont {Truhlar},\ and\ \citenamefont {Gagliardi}}]{scottAnalytic2021}%
  \BibitemOpen
  \bibfield  {author} {\bibinfo {author} {\bibfnamefont {T.~R.}\ \bibnamefont {Scott}}, \bibinfo {author} {\bibfnamefont {M.~S.}\ \bibnamefont {Oakley}}, \bibinfo {author} {\bibfnamefont {M.~R.}\ \bibnamefont {Hermes}}, \bibinfo {author} {\bibfnamefont {A.~M.}\ \bibnamefont {Sand}}, \bibinfo {author} {\bibfnamefont {R.}~\bibnamefont {Lindh}}, \bibinfo {author} {\bibfnamefont {D.~G.}\ \bibnamefont {Truhlar}},\ and\ \bibinfo {author} {\bibfnamefont {L.}~\bibnamefont {Gagliardi}},\ }\href {https://doi.org/10.1063/5.0039258} {\bibfield  {journal} {\bibinfo  {journal} {J. Chem. Phys.}\ }\textbf {\bibinfo {volume} {154}},\ \bibinfo {pages} {074108} (\bibinfo {year} {2021})}\BibitemShut {NoStop}%
\bibitem [{\citenamefont {Scott}\ \emph {et~al.}(2020)\citenamefont {Scott}, \citenamefont {Hermes}, \citenamefont {Sand}, \citenamefont {Oakley}, \citenamefont {Truhlar},\ and\ \citenamefont {Gagliardi}}]{ScottAnalytic2020}%
  \BibitemOpen
  \bibfield  {author} {\bibinfo {author} {\bibfnamefont {T.~R.}\ \bibnamefont {Scott}}, \bibinfo {author} {\bibfnamefont {M.~R.}\ \bibnamefont {Hermes}}, \bibinfo {author} {\bibfnamefont {A.~M.}\ \bibnamefont {Sand}}, \bibinfo {author} {\bibfnamefont {M.~S.}\ \bibnamefont {Oakley}}, \bibinfo {author} {\bibfnamefont {D.~G.}\ \bibnamefont {Truhlar}},\ and\ \bibinfo {author} {\bibfnamefont {L.}~\bibnamefont {Gagliardi}},\ }\href {https://doi.org/10.1063/5.0007040} {\bibfield  {journal} {\bibinfo  {journal} {J. Chem. Phys.}\ }\textbf {\bibinfo {volume} {153}},\ \bibinfo {pages} {014106} (\bibinfo {year} {2020})}\BibitemShut {NoStop}%
\bibitem [{\citenamefont {Sand}\ \emph {et~al.}(2018)\citenamefont {Sand}, \citenamefont {Hoyer}, \citenamefont {Sharkas}, \citenamefont {Kidder}, \citenamefont {Lindh}, \citenamefont {Truhlar},\ and\ \citenamefont {Gagliardi}}]{SandAnalytic2018}%
  \BibitemOpen
  \bibfield  {author} {\bibinfo {author} {\bibfnamefont {A.~M.}\ \bibnamefont {Sand}}, \bibinfo {author} {\bibfnamefont {C.~E.}\ \bibnamefont {Hoyer}}, \bibinfo {author} {\bibfnamefont {K.}~\bibnamefont {Sharkas}}, \bibinfo {author} {\bibfnamefont {K.~M.}\ \bibnamefont {Kidder}}, \bibinfo {author} {\bibfnamefont {R.}~\bibnamefont {Lindh}}, \bibinfo {author} {\bibfnamefont {D.~G.}\ \bibnamefont {Truhlar}},\ and\ \bibinfo {author} {\bibfnamefont {L.}~\bibnamefont {Gagliardi}},\ }\href {https://doi.org/10.1021/acs.jctc.7b00967} {\bibfield  {journal} {\bibinfo  {journal} {J. Chem. Theory Comput.}\ }\textbf {\bibinfo {volume} {14}},\ \bibinfo {pages} {126} (\bibinfo {year} {2018})}\BibitemShut {NoStop}%
\bibitem [{\citenamefont {Yang}\ \emph {et~al.}(2022)\citenamefont {Yang}, \citenamefont {Bonati}, \citenamefont {Polino},\ and\ \citenamefont {Parrinello}}]{yang2022using}%
  \BibitemOpen
  \bibfield  {author} {\bibinfo {author} {\bibfnamefont {M.}~\bibnamefont {Yang}}, \bibinfo {author} {\bibfnamefont {L.}~\bibnamefont {Bonati}}, \bibinfo {author} {\bibfnamefont {D.}~\bibnamefont {Polino}},\ and\ \bibinfo {author} {\bibfnamefont {M.}~\bibnamefont {Parrinello}},\ }\href@noop {} {\bibfield  {journal} {\bibinfo  {journal} {Catal. Today}\ }\textbf {\bibinfo {volume} {387}},\ \bibinfo {pages} {143} (\bibinfo {year} {2022})}\BibitemShut {NoStop}%
\bibitem [{\citenamefont {David}\ \emph {et~al.}(2025)\citenamefont {David}, \citenamefont {Puente}, \citenamefont {Gomez}, \citenamefont {Anton}, \citenamefont {Stirnemann},\ and\ \citenamefont {Laage}}]{david2025arcann}%
  \BibitemOpen
  \bibfield  {author} {\bibinfo {author} {\bibfnamefont {R.}~\bibnamefont {David}}, \bibinfo {author} {\bibfnamefont {M.~d.~l.}\ \bibnamefont {Puente}}, \bibinfo {author} {\bibfnamefont {A.}~\bibnamefont {Gomez}}, \bibinfo {author} {\bibfnamefont {O.}~\bibnamefont {Anton}}, \bibinfo {author} {\bibfnamefont {G.}~\bibnamefont {Stirnemann}},\ and\ \bibinfo {author} {\bibfnamefont {D.}~\bibnamefont {Laage}},\ }\href {https://doi.org/10.1039/D4DD00209A} {\bibfield  {journal} {\bibinfo  {journal} {Digit. Discov.}\ }\textbf {\bibinfo {volume} {4}},\ \bibinfo {pages} {54} (\bibinfo {year} {2025})}\BibitemShut {NoStop}%
\bibitem [{\citenamefont {Laio}\ and\ \citenamefont {Parrinello}(2002)}]{laio_escaping_2002}%
  \BibitemOpen
  \bibfield  {author} {\bibinfo {author} {\bibfnamefont {A.}~\bibnamefont {Laio}}\ and\ \bibinfo {author} {\bibfnamefont {M.}~\bibnamefont {Parrinello}},\ }\href {https://doi.org/10.1073/pnas.202427399} {\bibfield  {journal} {\bibinfo  {journal} {Proc. Natl. Acad. Sci. U. S. A.}\ }\textbf {\bibinfo {volume} {99}},\ \bibinfo {pages} {12562} (\bibinfo {year} {2002})}\BibitemShut {NoStop}%
\bibitem [{\citenamefont {Invernizzi}\ and\ \citenamefont {Parrinello}(2020)}]{invernizzi_rethinking_2020}%
  \BibitemOpen
  \bibfield  {author} {\bibinfo {author} {\bibfnamefont {M.}~\bibnamefont {Invernizzi}}\ and\ \bibinfo {author} {\bibfnamefont {M.}~\bibnamefont {Parrinello}},\ }\href {https://doi.org/10.1021/acs.jpclett.0c00497} {\bibfield  {journal} {\bibinfo  {journal} {J. Phys. Chem. Lett.}\ }\textbf {\bibinfo {volume} {11}},\ \bibinfo {pages} {2731} (\bibinfo {year} {2020})}\BibitemShut {NoStop}%
\bibitem [{\citenamefont {Trizio}\ \emph {et~al.}(2024)\citenamefont {Trizio}, \citenamefont {Rizzi}, \citenamefont {Piaggi}, \citenamefont {Invernizzi},\ and\ \citenamefont {Bonati}}]{trizio2024advanced}%
  \BibitemOpen
  \bibfield  {author} {\bibinfo {author} {\bibfnamefont {E.}~\bibnamefont {Trizio}}, \bibinfo {author} {\bibfnamefont {A.}~\bibnamefont {Rizzi}}, \bibinfo {author} {\bibfnamefont {P.~M.}\ \bibnamefont {Piaggi}}, \bibinfo {author} {\bibfnamefont {M.}~\bibnamefont {Invernizzi}},\ and\ \bibinfo {author} {\bibfnamefont {L.}~\bibnamefont {Bonati}},\ }\href {https://doi.org/10.48550/arXiv.2410.18019} {\bibinfo {title} {Advanced simulations with {PLUMED}: {OPES} and {Machine} {Learning} {Collective} {Variables}}} (\bibinfo {year} {2024}),\ \bibinfo {note} {arXiv:2410.18019 [physics]}\BibitemShut {NoStop}%
\bibitem [{\citenamefont {Smith}\ \emph {et~al.}(2018)\citenamefont {Smith}, \citenamefont {Nebgen}, \citenamefont {Lubbers}, \citenamefont {Isayev},\ and\ \citenamefont {Roitberg}}]{smith_less_2018}%
  \BibitemOpen
  \bibfield  {author} {\bibinfo {author} {\bibfnamefont {J.~S.}\ \bibnamefont {Smith}}, \bibinfo {author} {\bibfnamefont {B.}~\bibnamefont {Nebgen}}, \bibinfo {author} {\bibfnamefont {N.}~\bibnamefont {Lubbers}}, \bibinfo {author} {\bibfnamefont {O.}~\bibnamefont {Isayev}},\ and\ \bibinfo {author} {\bibfnamefont {A.~E.}\ \bibnamefont {Roitberg}},\ }\href {https://doi.org/10.1063/1.5023802} {\bibfield  {journal} {\bibinfo  {journal} {J. Chem. Phys.}\ }\textbf {\bibinfo {volume} {148}},\ \bibinfo {pages} {241733} (\bibinfo {year} {2018})}\BibitemShut {NoStop}%
\bibitem [{\citenamefont {Ang}\ \emph {et~al.}(2021)\citenamefont {Ang}, \citenamefont {Wang}, \citenamefont {Schwalbe-Koda}, \citenamefont {Axelrod},\ and\ \citenamefont {Gómez-Bombarelli}}]{ang_active_2021}%
  \BibitemOpen
  \bibfield  {author} {\bibinfo {author} {\bibfnamefont {S.~J.}\ \bibnamefont {Ang}}, \bibinfo {author} {\bibfnamefont {W.}~\bibnamefont {Wang}}, \bibinfo {author} {\bibfnamefont {D.}~\bibnamefont {Schwalbe-Koda}}, \bibinfo {author} {\bibfnamefont {S.}~\bibnamefont {Axelrod}},\ and\ \bibinfo {author} {\bibfnamefont {R.}~\bibnamefont {Gómez-Bombarelli}},\ }\href {https://doi.org/10.1016/j.chempr.2020.12.009} {\bibfield  {journal} {\bibinfo  {journal} {Chem}\ }\textbf {\bibinfo {volume} {7}},\ \bibinfo {pages} {738} (\bibinfo {year} {2021})}\BibitemShut {NoStop}%
\bibitem [{\citenamefont {Herr}\ \emph {et~al.}(2018)\citenamefont {Herr}, \citenamefont {Yao}, \citenamefont {McIntyre}, \citenamefont {Toth},\ and\ \citenamefont {Parkhill}}]{herr_metadynamics_2018}%
  \BibitemOpen
  \bibfield  {author} {\bibinfo {author} {\bibfnamefont {J.~E.}\ \bibnamefont {Herr}}, \bibinfo {author} {\bibfnamefont {K.}~\bibnamefont {Yao}}, \bibinfo {author} {\bibfnamefont {R.}~\bibnamefont {McIntyre}}, \bibinfo {author} {\bibfnamefont {D.~W.}\ \bibnamefont {Toth}},\ and\ \bibinfo {author} {\bibfnamefont {J.}~\bibnamefont {Parkhill}},\ }\href {https://doi.org/10.1063/1.5020067} {\bibfield  {journal} {\bibinfo  {journal} {J. Chem. Phys.}\ }\textbf {\bibinfo {volume} {148}},\ \bibinfo {pages} {241710} (\bibinfo {year} {2018})}\BibitemShut {NoStop}%
\bibitem [{\citenamefont {Benayad}\ \emph {et~al.}(2024)\citenamefont {Benayad}, \citenamefont {David},\ and\ \citenamefont {Stirnemann}}]{benayad_prebiotic_2024}%
  \BibitemOpen
  \bibfield  {author} {\bibinfo {author} {\bibfnamefont {Z.}~\bibnamefont {Benayad}}, \bibinfo {author} {\bibfnamefont {R.}~\bibnamefont {David}},\ and\ \bibinfo {author} {\bibfnamefont {G.}~\bibnamefont {Stirnemann}},\ }\href {https://doi.org/10.1073/pnas.2322040121} {\bibfield  {journal} {\bibinfo  {journal} {Proc. Natl. Acad. Sci. U. S. A.}\ }\textbf {\bibinfo {volume} {121}},\ \bibinfo {pages} {e2322040121} (\bibinfo {year} {2024})}\BibitemShut {NoStop}%
\bibitem [{\citenamefont {Perego}\ and\ \citenamefont {Bonati}(2024)}]{perego_data_2024}%
  \BibitemOpen
  \bibfield  {author} {\bibinfo {author} {\bibfnamefont {S.}~\bibnamefont {Perego}}\ and\ \bibinfo {author} {\bibfnamefont {L.}~\bibnamefont {Bonati}},\ }\href {https://doi.org/10.1038/s41524-024-01481-6} {\bibfield  {journal} {\bibinfo  {journal} {Npj Comput. Mater.}\ }\textbf {\bibinfo {volume} {10}},\ \bibinfo {pages} {1} (\bibinfo {year} {2024})}\BibitemShut {NoStop}%
\bibitem [{\citenamefont {Saade}\ and\ \citenamefont {Burton}(2024)}]{saade_excited_2024}%
  \BibitemOpen
  \bibfield  {author} {\bibinfo {author} {\bibfnamefont {S.}~\bibnamefont {Saade}}\ and\ \bibinfo {author} {\bibfnamefont {H.~G.~A.}\ \bibnamefont {Burton}},\ }\href {https://doi.org/10.1021/acs.jctc.4c00212} {\bibfield  {journal} {\bibinfo  {journal} {J. Chem. Theory Comput.}\ }\textbf {\bibinfo {volume} {20}},\ \bibinfo {pages} {5105} (\bibinfo {year} {2024})}\BibitemShut {NoStop}%
\bibitem [{\citenamefont {Khedkar}\ and\ \citenamefont {Roemelt}(2019)}]{khedkar_active_2019}%
  \BibitemOpen
  \bibfield  {author} {\bibinfo {author} {\bibfnamefont {A.}~\bibnamefont {Khedkar}}\ and\ \bibinfo {author} {\bibfnamefont {M.}~\bibnamefont {Roemelt}},\ }\href {https://doi.org/10.1021/acs.jctc.8b01293} {\bibfield  {journal} {\bibinfo  {journal} {J. Chem. Theory Comput.}\ }\textbf {\bibinfo {volume} {15}},\ \bibinfo {pages} {3522} (\bibinfo {year} {2019})}\BibitemShut {NoStop}%
\bibitem [{\citenamefont {Sayfutyarova}\ \emph {et~al.}(2017)\citenamefont {Sayfutyarova}, \citenamefont {Sun}, \citenamefont {Chan},\ and\ \citenamefont {Knizia}}]{sayfutyarova_automated_2017}%
  \BibitemOpen
  \bibfield  {author} {\bibinfo {author} {\bibfnamefont {E.~R.}\ \bibnamefont {Sayfutyarova}}, \bibinfo {author} {\bibfnamefont {Q.}~\bibnamefont {Sun}}, \bibinfo {author} {\bibfnamefont {G.~K.-L.}\ \bibnamefont {Chan}},\ and\ \bibinfo {author} {\bibfnamefont {G.}~\bibnamefont {Knizia}},\ }\href {https://doi.org/10.1021/acs.jctc.7b00128} {\bibfield  {journal} {\bibinfo  {journal} {J. Chem. Theory Comput.}\ }\textbf {\bibinfo {volume} {13}},\ \bibinfo {pages} {4063} (\bibinfo {year} {2017})}\BibitemShut {NoStop}%
\bibitem [{\citenamefont {Stein}\ and\ \citenamefont {Reiher}(2019)}]{stein_autocas_2019}%
  \BibitemOpen
  \bibfield  {author} {\bibinfo {author} {\bibfnamefont {C.~J.}\ \bibnamefont {Stein}}\ and\ \bibinfo {author} {\bibfnamefont {M.}~\bibnamefont {Reiher}},\ }\href {https://doi.org/10.1002/jcc.25869} {\bibfield  {journal} {\bibinfo  {journal} {J. Comput. Chem.}\ }\textbf {\bibinfo {volume} {40}},\ \bibinfo {pages} {2216} (\bibinfo {year} {2019})}\BibitemShut {NoStop}%
\bibitem [{\citenamefont {King}\ and\ \citenamefont {Gagliardi}(2021)}]{king_ranked-orbital_2021}%
  \BibitemOpen
  \bibfield  {author} {\bibinfo {author} {\bibfnamefont {D.~S.}\ \bibnamefont {King}}\ and\ \bibinfo {author} {\bibfnamefont {L.}~\bibnamefont {Gagliardi}},\ }\href {https://doi.org/10.1021/acs.jctc.1c00037} {\bibfield  {journal} {\bibinfo  {journal} {J. Chem. Theory Comput.}\ }\textbf {\bibinfo {volume} {17}},\ \bibinfo {pages} {2817} (\bibinfo {year} {2021})}\BibitemShut {NoStop}%
\bibitem [{\citenamefont {Jeong}\ \emph {et~al.}(2020)\citenamefont {Jeong}, \citenamefont {Stoneburner}, \citenamefont {King}, \citenamefont {Li}, \citenamefont {Walker}, \citenamefont {Lindh},\ and\ \citenamefont {Gagliardi}}]{jeong_automation_2020}%
  \BibitemOpen
  \bibfield  {author} {\bibinfo {author} {\bibfnamefont {W.}~\bibnamefont {Jeong}}, \bibinfo {author} {\bibfnamefont {S.~J.}\ \bibnamefont {Stoneburner}}, \bibinfo {author} {\bibfnamefont {D.}~\bibnamefont {King}}, \bibinfo {author} {\bibfnamefont {R.}~\bibnamefont {Li}}, \bibinfo {author} {\bibfnamefont {A.}~\bibnamefont {Walker}}, \bibinfo {author} {\bibfnamefont {R.}~\bibnamefont {Lindh}},\ and\ \bibinfo {author} {\bibfnamefont {L.}~\bibnamefont {Gagliardi}},\ }\href {https://doi.org/10.1021/acs.jctc.9b01297} {\bibfield  {journal} {\bibinfo  {journal} {J. Chem. Theory Comput.}\ }\textbf {\bibinfo {volume} {16}},\ \bibinfo {pages} {2389} (\bibinfo {year} {2020})}\BibitemShut {NoStop}%
\bibitem [{\citenamefont {Wen}\ \emph {et~al.}(2024)\citenamefont {Wen}, \citenamefont {Boyn}, \citenamefont {Martirez}, \citenamefont {Zhao},\ and\ \citenamefont {Carter}}]{wen_strategies_2024}%
  \BibitemOpen
  \bibfield  {author} {\bibinfo {author} {\bibfnamefont {X.}~\bibnamefont {Wen}}, \bibinfo {author} {\bibfnamefont {J.-N.}\ \bibnamefont {Boyn}}, \bibinfo {author} {\bibfnamefont {J.~M.~P.}\ \bibnamefont {Martirez}}, \bibinfo {author} {\bibfnamefont {Q.}~\bibnamefont {Zhao}},\ and\ \bibinfo {author} {\bibfnamefont {E.~A.}\ \bibnamefont {Carter}},\ }\href {https://doi.org/10.1021/acs.jctc.4c00558} {\bibfield  {journal} {\bibinfo  {journal} {J. Chem. Theory Comput.}\ }\textbf {\bibinfo {volume} {20}},\ \bibinfo {pages} {6037} (\bibinfo {year} {2024})}\BibitemShut {NoStop}%
\bibitem [{\citenamefont {Han}\ and\ \citenamefont {Luber}(2020)}]{han_complete_2020}%
  \BibitemOpen
  \bibfield  {author} {\bibinfo {author} {\bibfnamefont {R.}~\bibnamefont {Han}}\ and\ \bibinfo {author} {\bibfnamefont {S.}~\bibnamefont {Luber}},\ }\href {https://doi.org/10.1002/jcc.26201} {\bibfield  {journal} {\bibinfo  {journal} {J. Comput. Chem.}\ }\textbf {\bibinfo {volume} {41}},\ \bibinfo {pages} {1586} (\bibinfo {year} {2020})}\BibitemShut {NoStop}%
\bibitem [{\citenamefont {Geng}\ \emph {et~al.}(2019)\citenamefont {Geng}, \citenamefont {Weiske}, \citenamefont {Li}, \citenamefont {Shaik},\ and\ \citenamefont {Schwarz}}]{geng_intrinsic_2019}%
  \BibitemOpen
  \bibfield  {author} {\bibinfo {author} {\bibfnamefont {C.}~\bibnamefont {Geng}}, \bibinfo {author} {\bibfnamefont {T.}~\bibnamefont {Weiske}}, \bibinfo {author} {\bibfnamefont {J.}~\bibnamefont {Li}}, \bibinfo {author} {\bibfnamefont {S.}~\bibnamefont {Shaik}},\ and\ \bibinfo {author} {\bibfnamefont {H.}~\bibnamefont {Schwarz}},\ }\href {https://doi.org/10.1021/jacs.8b11739} {\bibfield  {journal} {\bibinfo  {journal} {J. Am. Chem. Soc.}\ }\textbf {\bibinfo {volume} {141}},\ \bibinfo {pages} {599} (\bibinfo {year} {2019})}\BibitemShut {NoStop}%
\bibitem [{\citenamefont {Batatia}\ \emph {et~al.}(2022)\citenamefont {Batatia}, \citenamefont {Kovacs}, \citenamefont {Simm}, \citenamefont {Ortner},\ and\ \citenamefont {Csanyi}}]{Batatia2022MACE:Fields}%
  \BibitemOpen
  \bibfield  {author} {\bibinfo {author} {\bibfnamefont {I.}~\bibnamefont {Batatia}}, \bibinfo {author} {\bibfnamefont {D.~P.}\ \bibnamefont {Kovacs}}, \bibinfo {author} {\bibfnamefont {G.}~\bibnamefont {Simm}}, \bibinfo {author} {\bibfnamefont {C.}~\bibnamefont {Ortner}},\ and\ \bibinfo {author} {\bibfnamefont {G.}~\bibnamefont {Csanyi}},\ }\href {https://proceedings.neurips.cc/paper_files/paper/2022/hash/4a36c3c51af11ed9f34615b81edb5bbc-Abstract-Conference.html} {\bibfield  {journal} {\bibinfo  {journal} {Adv. Neural Inf. Process Syst.}\ }\textbf {\bibinfo {volume} {35}},\ \bibinfo {pages} {11423} (\bibinfo {year} {2022})}\BibitemShut {NoStop}%
\bibitem [{\citenamefont {Owen}\ \emph {et~al.}(2024)\citenamefont {Owen}, \citenamefont {Torrisi}, \citenamefont {Xie}, \citenamefont {Batzner}, \citenamefont {Bystrom}, \citenamefont {Coulter}, \citenamefont {Musaelian}, \citenamefont {Sun},\ and\ \citenamefont {Kozinsky}}]{owen_complexity_2024}%
  \BibitemOpen
  \bibfield  {author} {\bibinfo {author} {\bibfnamefont {C.~J.}\ \bibnamefont {Owen}}, \bibinfo {author} {\bibfnamefont {S.~B.}\ \bibnamefont {Torrisi}}, \bibinfo {author} {\bibfnamefont {Y.}~\bibnamefont {Xie}}, \bibinfo {author} {\bibfnamefont {S.}~\bibnamefont {Batzner}}, \bibinfo {author} {\bibfnamefont {K.}~\bibnamefont {Bystrom}}, \bibinfo {author} {\bibfnamefont {J.}~\bibnamefont {Coulter}}, \bibinfo {author} {\bibfnamefont {A.}~\bibnamefont {Musaelian}}, \bibinfo {author} {\bibfnamefont {L.}~\bibnamefont {Sun}},\ and\ \bibinfo {author} {\bibfnamefont {B.}~\bibnamefont {Kozinsky}},\ }\href {https://doi.org/10.1038/s41524-024-01264-z} {\bibfield  {journal} {\bibinfo  {journal} {Npj Comput. Mater.}\ }\textbf {\bibinfo {volume} {10}},\ \bibinfo {pages} {1} (\bibinfo {year} {2024})}\BibitemShut {NoStop}%
\bibitem [{\citenamefont {Marie}\ and\ \citenamefont {Burton}(2023)}]{marie_excited_2023}%
  \BibitemOpen
  \bibfield  {author} {\bibinfo {author} {\bibfnamefont {A.}~\bibnamefont {Marie}}\ and\ \bibinfo {author} {\bibfnamefont {H.~G.~A.}\ \bibnamefont {Burton}},\ }\href {https://doi.org/10.1021/acs.jpca.3c00603} {\bibfield  {journal} {\bibinfo  {journal} {J. Phys. Chem. A}\ }\textbf {\bibinfo {volume} {127}},\ \bibinfo {pages} {4538} (\bibinfo {year} {2023})}\BibitemShut {NoStop}%
\bibitem [{\citenamefont {Eyring}(1935)}]{eyring_activated_1935}%
  \BibitemOpen
  \bibfield  {author} {\bibinfo {author} {\bibfnamefont {H.}~\bibnamefont {Eyring}},\ }\href {https://doi.org/10.1063/1.1749604} {\bibfield  {journal} {\bibinfo  {journal} {J. Chem. Phys.}\ }\textbf {\bibinfo {volume} {3}},\ \bibinfo {pages} {107} (\bibinfo {year} {1935})}\BibitemShut {NoStop}%
\bibitem [{\citenamefont {Novelli}\ \emph {et~al.}(2025)\citenamefont {Novelli}, \citenamefont {Meanti}, \citenamefont {Buigues}, \citenamefont {Rosasco}, \citenamefont {Parrinello}, \citenamefont {Pontil},\ and\ \citenamefont {Bonati}}]{franken}%
  \BibitemOpen
  \bibfield  {author} {\bibinfo {author} {\bibfnamefont {P.}~\bibnamefont {Novelli}}, \bibinfo {author} {\bibfnamefont {G.}~\bibnamefont {Meanti}}, \bibinfo {author} {\bibfnamefont {P.~J.}\ \bibnamefont {Buigues}}, \bibinfo {author} {\bibfnamefont {L.}~\bibnamefont {Rosasco}}, \bibinfo {author} {\bibfnamefont {M.}~\bibnamefont {Parrinello}}, \bibinfo {author} {\bibfnamefont {M.}~\bibnamefont {Pontil}},\ and\ \bibinfo {author} {\bibfnamefont {L.}~\bibnamefont {Bonati}},\ }\href {https://doi.org/10.48550/arXiv.2505.05652} {\bibinfo {title} {Fast and {Fourier} {Features} for {Transfer} {Learning} of {Interatomic} {Potentials}}} (\bibinfo {year} {2025}),\ \bibinfo {note} {arXiv:2505.05652 [physics]}\BibitemShut {NoStop}%
\bibitem [{\citenamefont {Chen}\ \emph {et~al.}(2023)\citenamefont {Chen}, \citenamefont {Lee}, \citenamefont {Ye}, \citenamefont {Berkelbach}, \citenamefont {Reichman},\ and\ \citenamefont {Markland}}]{chen2023data}%
  \BibitemOpen
  \bibfield  {author} {\bibinfo {author} {\bibfnamefont {M.~S.}\ \bibnamefont {Chen}}, \bibinfo {author} {\bibfnamefont {J.}~\bibnamefont {Lee}}, \bibinfo {author} {\bibfnamefont {H.-Z.}\ \bibnamefont {Ye}}, \bibinfo {author} {\bibfnamefont {T.~C.}\ \bibnamefont {Berkelbach}}, \bibinfo {author} {\bibfnamefont {D.~R.}\ \bibnamefont {Reichman}},\ and\ \bibinfo {author} {\bibfnamefont {T.~E.}\ \bibnamefont {Markland}},\ }\href {https://doi.org/10.1021/acs.jctc.2c01203} {\bibfield  {journal} {\bibinfo  {journal} {J. Chem. Theory Comput.}\ }\textbf {\bibinfo {volume} {19}},\ \bibinfo {pages} {4510} (\bibinfo {year} {2023})}\BibitemShut {NoStop}%
\bibitem [{\citenamefont {Bocus}\ \emph {et~al.}(2025)\citenamefont {Bocus}, \citenamefont {Vandenhaute},\ and\ \citenamefont {Van~Speybroeck}}]{bocus2025operando}%
  \BibitemOpen
  \bibfield  {author} {\bibinfo {author} {\bibfnamefont {M.}~\bibnamefont {Bocus}}, \bibinfo {author} {\bibfnamefont {S.}~\bibnamefont {Vandenhaute}},\ and\ \bibinfo {author} {\bibfnamefont {V.}~\bibnamefont {Van~Speybroeck}},\ }\href {https://doi.org/10.1002/anie.202413637} {\bibfield  {journal} {\bibinfo  {journal} {Angew. Chem., Int. Ed. Engl.}\ }\textbf {\bibinfo {volume} {64}},\ \bibinfo {pages} {e202413637} (\bibinfo {year} {2025})}\BibitemShut {NoStop}%
\bibitem [{\citenamefont {Sun}\ \emph {et~al.}(2018)\citenamefont {Sun}, \citenamefont {Berkelbach}, \citenamefont {Blunt}, \citenamefont {Booth}, \citenamefont {Guo}, \citenamefont {Li}, \citenamefont {Liu}, \citenamefont {McClain}, \citenamefont {Sayfutyarova}, \citenamefont {Sharma}, \citenamefont {Wouters},\ and\ \citenamefont {Chan}}]{sunPySCF2018}%
  \BibitemOpen
  \bibfield  {author} {\bibinfo {author} {\bibfnamefont {Q.}~\bibnamefont {Sun}}, \bibinfo {author} {\bibfnamefont {T.~C.}\ \bibnamefont {Berkelbach}}, \bibinfo {author} {\bibfnamefont {N.~S.}\ \bibnamefont {Blunt}}, \bibinfo {author} {\bibfnamefont {G.~H.}\ \bibnamefont {Booth}}, \bibinfo {author} {\bibfnamefont {S.}~\bibnamefont {Guo}}, \bibinfo {author} {\bibfnamefont {Z.}~\bibnamefont {Li}}, \bibinfo {author} {\bibfnamefont {J.}~\bibnamefont {Liu}}, \bibinfo {author} {\bibfnamefont {J.~D.}\ \bibnamefont {McClain}}, \bibinfo {author} {\bibfnamefont {E.~R.}\ \bibnamefont {Sayfutyarova}}, \bibinfo {author} {\bibfnamefont {S.}~\bibnamefont {Sharma}}, \bibinfo {author} {\bibfnamefont {S.}~\bibnamefont {Wouters}},\ and\ \bibinfo {author} {\bibfnamefont {G.~K.-L.}\ \bibnamefont {Chan}},\ }\href {https://doi.org/10.1002/wcms.1340} {\bibfield  {journal} {\bibinfo  {journal} {Wiley Interdiscip. Rev. Comput. Mol. Sci.}\ }\textbf {\bibinfo {volume} {8}},\ \bibinfo {pages} {e1340} (\bibinfo {year} {2018})}\BibitemShut
  {NoStop}%
\bibitem [{\citenamefont {Sun}\ \emph {et~al.}(2020)\citenamefont {Sun}, \citenamefont {Zhang}, \citenamefont {Banerjee}, \citenamefont {Bao}, \citenamefont {Barbry}, \citenamefont {Blunt}, \citenamefont {Bogdanov}, \citenamefont {Booth}, \citenamefont {Chen}, \citenamefont {Cui}, \citenamefont {Eriksen}, \citenamefont {Gao}, \citenamefont {Guo}, \citenamefont {Hermann}, \citenamefont {Hermes}, \citenamefont {Koh}, \citenamefont {Koval}, \citenamefont {Lehtola}, \citenamefont {Li}, \citenamefont {Liu}, \citenamefont {Mardirossian}, \citenamefont {McClain}, \citenamefont {Motta}, \citenamefont {Mussard}, \citenamefont {Pham}, \citenamefont {Pulkin}, \citenamefont {Purwanto}, \citenamefont {Robinson}, \citenamefont {Ronca}, \citenamefont {Sayfutyarova}, \citenamefont {Scheurer}, \citenamefont {Schurkus}, \citenamefont {Smith}, \citenamefont {Sun}, \citenamefont {Sun}, \citenamefont {Upadhyay}, \citenamefont {Wagner}, \citenamefont {Wang}, \citenamefont {White}, \citenamefont {Whitfield}, \citenamefont
  {Williamson}, \citenamefont {Wouters}, \citenamefont {Yang}, \citenamefont {Yu}, \citenamefont {Zhu}, \citenamefont {Berkelbach}, \citenamefont {Sharma}, \citenamefont {Sokolov},\ and\ \citenamefont {Chan}}]{sunRecent2020}%
  \BibitemOpen
  \bibfield  {author} {\bibinfo {author} {\bibfnamefont {Q.}~\bibnamefont {Sun}}, \bibinfo {author} {\bibfnamefont {X.}~\bibnamefont {Zhang}}, \bibinfo {author} {\bibfnamefont {S.}~\bibnamefont {Banerjee}}, \bibinfo {author} {\bibfnamefont {P.}~\bibnamefont {Bao}}, \bibinfo {author} {\bibfnamefont {M.}~\bibnamefont {Barbry}}, \bibinfo {author} {\bibfnamefont {N.~S.}\ \bibnamefont {Blunt}}, \bibinfo {author} {\bibfnamefont {N.~A.}\ \bibnamefont {Bogdanov}}, \bibinfo {author} {\bibfnamefont {G.~H.}\ \bibnamefont {Booth}}, \bibinfo {author} {\bibfnamefont {J.}~\bibnamefont {Chen}}, \bibinfo {author} {\bibfnamefont {Z.-H.}\ \bibnamefont {Cui}}, \bibinfo {author} {\bibfnamefont {J.~J.}\ \bibnamefont {Eriksen}}, \bibinfo {author} {\bibfnamefont {Y.}~\bibnamefont {Gao}}, \bibinfo {author} {\bibfnamefont {S.}~\bibnamefont {Guo}}, \bibinfo {author} {\bibfnamefont {J.}~\bibnamefont {Hermann}}, \bibinfo {author} {\bibfnamefont {M.~R.}\ \bibnamefont {Hermes}}, \bibinfo {author} {\bibfnamefont {K.}~\bibnamefont {Koh}},
  \bibinfo {author} {\bibfnamefont {P.}~\bibnamefont {Koval}}, \bibinfo {author} {\bibfnamefont {S.}~\bibnamefont {Lehtola}}, \bibinfo {author} {\bibfnamefont {Z.}~\bibnamefont {Li}}, \bibinfo {author} {\bibfnamefont {J.}~\bibnamefont {Liu}}, \bibinfo {author} {\bibfnamefont {N.}~\bibnamefont {Mardirossian}}, \bibinfo {author} {\bibfnamefont {J.~D.}\ \bibnamefont {McClain}}, \bibinfo {author} {\bibfnamefont {M.}~\bibnamefont {Motta}}, \bibinfo {author} {\bibfnamefont {B.}~\bibnamefont {Mussard}}, \bibinfo {author} {\bibfnamefont {H.~Q.}\ \bibnamefont {Pham}}, \bibinfo {author} {\bibfnamefont {A.}~\bibnamefont {Pulkin}}, \bibinfo {author} {\bibfnamefont {W.}~\bibnamefont {Purwanto}}, \bibinfo {author} {\bibfnamefont {P.~J.}\ \bibnamefont {Robinson}}, \bibinfo {author} {\bibfnamefont {E.}~\bibnamefont {Ronca}}, \bibinfo {author} {\bibfnamefont {E.~R.}\ \bibnamefont {Sayfutyarova}}, \bibinfo {author} {\bibfnamefont {M.}~\bibnamefont {Scheurer}}, \bibinfo {author} {\bibfnamefont {H.~F.}\ \bibnamefont {Schurkus}},
  \bibinfo {author} {\bibfnamefont {J.~E.~T.}\ \bibnamefont {Smith}}, \bibinfo {author} {\bibfnamefont {C.}~\bibnamefont {Sun}}, \bibinfo {author} {\bibfnamefont {S.-N.}\ \bibnamefont {Sun}}, \bibinfo {author} {\bibfnamefont {S.}~\bibnamefont {Upadhyay}}, \bibinfo {author} {\bibfnamefont {L.~K.}\ \bibnamefont {Wagner}}, \bibinfo {author} {\bibfnamefont {X.}~\bibnamefont {Wang}}, \bibinfo {author} {\bibfnamefont {A.}~\bibnamefont {White}}, \bibinfo {author} {\bibfnamefont {J.~D.}\ \bibnamefont {Whitfield}}, \bibinfo {author} {\bibfnamefont {M.~J.}\ \bibnamefont {Williamson}}, \bibinfo {author} {\bibfnamefont {S.}~\bibnamefont {Wouters}}, \bibinfo {author} {\bibfnamefont {J.}~\bibnamefont {Yang}}, \bibinfo {author} {\bibfnamefont {J.~M.}\ \bibnamefont {Yu}}, \bibinfo {author} {\bibfnamefont {T.}~\bibnamefont {Zhu}}, \bibinfo {author} {\bibfnamefont {T.~C.}\ \bibnamefont {Berkelbach}}, \bibinfo {author} {\bibfnamefont {S.}~\bibnamefont {Sharma}}, \bibinfo {author} {\bibfnamefont {A.~Y.}\ \bibnamefont
  {Sokolov}},\ and\ \bibinfo {author} {\bibfnamefont {G.~K.-L.}\ \bibnamefont {Chan}},\ }\href {https://doi.org/10.1063/5.0006074} {\bibfield  {journal} {\bibinfo  {journal} {J. Chem. Phys.}\ }\textbf {\bibinfo {volume} {153}},\ \bibinfo {pages} {024109} (\bibinfo {year} {2020})}\BibitemShut {NoStop}%
\bibitem [{\citenamefont {Sun}(2015)}]{SunLibcint2015}%
  \BibitemOpen
  \bibfield  {author} {\bibinfo {author} {\bibfnamefont {Q.}~\bibnamefont {Sun}},\ }\href {https://doi.org/10.1002/jcc.23981} {\bibfield  {journal} {\bibinfo  {journal} {J. Comput. Chem.}\ }\textbf {\bibinfo {volume} {36}},\ \bibinfo {pages} {1664} (\bibinfo {year} {2015})}\BibitemShut {NoStop}%
\bibitem [{\citenamefont {Marques}\ \emph {et~al.}(2012)\citenamefont {Marques}, \citenamefont {Oliveira},\ and\ \citenamefont {Burnus}}]{MarquesLibxc2012}%
  \BibitemOpen
  \bibfield  {author} {\bibinfo {author} {\bibfnamefont {M.~A.~L.}\ \bibnamefont {Marques}}, \bibinfo {author} {\bibfnamefont {M.~J.~T.}\ \bibnamefont {Oliveira}},\ and\ \bibinfo {author} {\bibfnamefont {T.}~\bibnamefont {Burnus}},\ }\href {https://doi.org/10.1016/j.cpc.2012.05.007} {\bibfield  {journal} {\bibinfo  {journal} {Comput. Phys. Commun.}\ }\textbf {\bibinfo {volume} {183}},\ \bibinfo {pages} {2272} (\bibinfo {year} {2012})}\BibitemShut {NoStop}%
\bibitem [{\citenamefont {Lehtola}\ \emph {et~al.}(2018)\citenamefont {Lehtola}, \citenamefont {Steigemann}, \citenamefont {Oliveira},\ and\ \citenamefont {Marques}}]{LehtolaRecent2018}%
  \BibitemOpen
  \bibfield  {author} {\bibinfo {author} {\bibfnamefont {S.}~\bibnamefont {Lehtola}}, \bibinfo {author} {\bibfnamefont {C.}~\bibnamefont {Steigemann}}, \bibinfo {author} {\bibfnamefont {M.~J.~T.}\ \bibnamefont {Oliveira}},\ and\ \bibinfo {author} {\bibfnamefont {M.~A.~L.}\ \bibnamefont {Marques}},\ }\href {https://doi.org/10.1016/j.softx.2017.11.002} {\bibfield  {journal} {\bibinfo  {journal} {SoftwareX}\ }\textbf {\bibinfo {volume} {7}},\ \bibinfo {pages} {1} (\bibinfo {year} {2018})}\BibitemShut {NoStop}%
\bibitem [{PyS(2025)}]{PySCFforge2025}%
  \BibitemOpen
  \href@noop {} {\bibinfo {title} {{{PySCF-forge}}}} (\bibinfo {year} {2025})\BibitemShut {NoStop}%
\bibitem [{\citenamefont {Wang}\ and\ \citenamefont {Song}(2016)}]{wangGeometry2016}%
  \BibitemOpen
  \bibfield  {author} {\bibinfo {author} {\bibfnamefont {L.-P.}\ \bibnamefont {Wang}}\ and\ \bibinfo {author} {\bibfnamefont {C.}~\bibnamefont {Song}},\ }\href {https://doi.org/10.1063/1.4952956} {\bibfield  {journal} {\bibinfo  {journal} {J. Chem. Phys.}\ }\textbf {\bibinfo {volume} {144}},\ \bibinfo {pages} {214108} (\bibinfo {year} {2016})}\BibitemShut {NoStop}%
\bibitem [{\citenamefont {Weigend}\ and\ \citenamefont {Ahlrichs}(2005)}]{def2tzvp}%
  \BibitemOpen
  \bibfield  {author} {\bibinfo {author} {\bibfnamefont {F.}~\bibnamefont {Weigend}}\ and\ \bibinfo {author} {\bibfnamefont {R.}~\bibnamefont {Ahlrichs}},\ }\href {https://doi.org/10.1039/B508541A} {\bibfield  {journal} {\bibinfo  {journal} {Phys. Chem. Chem. Phys.}\ }\textbf {\bibinfo {volume} {7}},\ \bibinfo {pages} {3297} (\bibinfo {year} {2005})}\BibitemShut {NoStop}%
\bibitem [{\citenamefont {Hjorth~Larsen}\ \emph {et~al.}(2017)\citenamefont {Hjorth~Larsen}, \citenamefont {Jørgen~Mortensen}, \citenamefont {Blomqvist}, \citenamefont {Castelli}, \citenamefont {Christensen}, \citenamefont {Dułak}, \citenamefont {Friis}, \citenamefont {Groves}, \citenamefont {Hammer}, \citenamefont {Hargus}, \citenamefont {Hermes}, \citenamefont {Jennings}, \citenamefont {Bjerre~Jensen}, \citenamefont {Kermode}, \citenamefont {Kitchin}, \citenamefont {Leonhard~Kolsbjerg}, \citenamefont {Kubal}, \citenamefont {Kaasbjerg}, \citenamefont {Lysgaard}, \citenamefont {Bergmann~Maronsson}, \citenamefont {Maxson}, \citenamefont {Olsen}, \citenamefont {Pastewka}, \citenamefont {Peterson}, \citenamefont {Rostgaard}, \citenamefont {Schiøtz}, \citenamefont {Schütt}, \citenamefont {Strange}, \citenamefont {Thygesen}, \citenamefont {Vegge}, \citenamefont {Vilhelmsen}, \citenamefont {Walter}, \citenamefont {Zeng},\ and\ \citenamefont {Jacobsen}}]{ase_2017}%
  \BibitemOpen
  \bibfield  {author} {\bibinfo {author} {\bibfnamefont {A.}~\bibnamefont {Hjorth~Larsen}}, \bibinfo {author} {\bibfnamefont {J.}~\bibnamefont {Jørgen~Mortensen}}, \bibinfo {author} {\bibfnamefont {J.}~\bibnamefont {Blomqvist}}, \bibinfo {author} {\bibfnamefont {I.~E.}\ \bibnamefont {Castelli}}, \bibinfo {author} {\bibfnamefont {R.}~\bibnamefont {Christensen}}, \bibinfo {author} {\bibfnamefont {M.}~\bibnamefont {Dułak}}, \bibinfo {author} {\bibfnamefont {J.}~\bibnamefont {Friis}}, \bibinfo {author} {\bibfnamefont {M.~N.}\ \bibnamefont {Groves}}, \bibinfo {author} {\bibfnamefont {B.}~\bibnamefont {Hammer}}, \bibinfo {author} {\bibfnamefont {C.}~\bibnamefont {Hargus}}, \bibinfo {author} {\bibfnamefont {E.~D.}\ \bibnamefont {Hermes}}, \bibinfo {author} {\bibfnamefont {P.~C.}\ \bibnamefont {Jennings}}, \bibinfo {author} {\bibfnamefont {P.}~\bibnamefont {Bjerre~Jensen}}, \bibinfo {author} {\bibfnamefont {J.}~\bibnamefont {Kermode}}, \bibinfo {author} {\bibfnamefont {J.~R.}\ \bibnamefont {Kitchin}}, \bibinfo
  {author} {\bibfnamefont {E.}~\bibnamefont {Leonhard~Kolsbjerg}}, \bibinfo {author} {\bibfnamefont {J.}~\bibnamefont {Kubal}}, \bibinfo {author} {\bibfnamefont {K.}~\bibnamefont {Kaasbjerg}}, \bibinfo {author} {\bibfnamefont {S.}~\bibnamefont {Lysgaard}}, \bibinfo {author} {\bibfnamefont {J.}~\bibnamefont {Bergmann~Maronsson}}, \bibinfo {author} {\bibfnamefont {T.}~\bibnamefont {Maxson}}, \bibinfo {author} {\bibfnamefont {T.}~\bibnamefont {Olsen}}, \bibinfo {author} {\bibfnamefont {L.}~\bibnamefont {Pastewka}}, \bibinfo {author} {\bibfnamefont {A.}~\bibnamefont {Peterson}}, \bibinfo {author} {\bibfnamefont {C.}~\bibnamefont {Rostgaard}}, \bibinfo {author} {\bibfnamefont {J.}~\bibnamefont {Schiøtz}}, \bibinfo {author} {\bibfnamefont {O.}~\bibnamefont {Schütt}}, \bibinfo {author} {\bibfnamefont {M.}~\bibnamefont {Strange}}, \bibinfo {author} {\bibfnamefont {K.~S.}\ \bibnamefont {Thygesen}}, \bibinfo {author} {\bibfnamefont {T.}~\bibnamefont {Vegge}}, \bibinfo {author} {\bibfnamefont {L.}~\bibnamefont
  {Vilhelmsen}}, \bibinfo {author} {\bibfnamefont {M.}~\bibnamefont {Walter}}, \bibinfo {author} {\bibfnamefont {Z.}~\bibnamefont {Zeng}},\ and\ \bibinfo {author} {\bibfnamefont {K.~W.}\ \bibnamefont {Jacobsen}},\ }\href {https://doi.org/10.1088/1361-648X/aa680e} {\bibfield  {journal} {\bibinfo  {journal} {J. Phys. Condens. Matter}\ }\textbf {\bibinfo {volume} {29}},\ \bibinfo {pages} {273002} (\bibinfo {year} {2017})}\BibitemShut {NoStop}%
\bibitem [{\citenamefont {Drautz}(2019)}]{Drautz2019AtomicPotentials}%
  \BibitemOpen
  \bibfield  {author} {\bibinfo {author} {\bibfnamefont {R.}~\bibnamefont {Drautz}},\ }\href {https://doi.org/10.1103/PHYSREVB.99.014104/FIGURES/7/MEDIUM} {\bibfield  {journal} {\bibinfo  {journal} {Phys. Rev. B}\ }\textbf {\bibinfo {volume} {99}},\ \bibinfo {pages} {014104} (\bibinfo {year} {2019})}\BibitemShut {NoStop}%
\bibitem [{\citenamefont {Vandermause}\ \emph {et~al.}(2020)\citenamefont {Vandermause}, \citenamefont {Torrisi}, \citenamefont {Batzner}, \citenamefont {Xie}, \citenamefont {Sun}, \citenamefont {Kolpak},\ and\ \citenamefont {Kozinsky}}]{Vandermause2020On-the-flyEvents}%
  \BibitemOpen
  \bibfield  {author} {\bibinfo {author} {\bibfnamefont {J.}~\bibnamefont {Vandermause}}, \bibinfo {author} {\bibfnamefont {S.~B.}\ \bibnamefont {Torrisi}}, \bibinfo {author} {\bibfnamefont {S.}~\bibnamefont {Batzner}}, \bibinfo {author} {\bibfnamefont {Y.}~\bibnamefont {Xie}}, \bibinfo {author} {\bibfnamefont {L.}~\bibnamefont {Sun}}, \bibinfo {author} {\bibfnamefont {A.~M.}\ \bibnamefont {Kolpak}},\ and\ \bibinfo {author} {\bibfnamefont {B.}~\bibnamefont {Kozinsky}},\ }\href {https://doi.org/10.1038/s41524-020-0283-z} {\bibfield  {journal} {\bibinfo  {journal} {Npj Comput. Mater.}\ }\textbf {\bibinfo {volume} {6}},\ \bibinfo {pages} {1} (\bibinfo {year} {2020})}\BibitemShut {NoStop}%
\bibitem [{\citenamefont {Thompson}\ \emph {et~al.}(2022)\citenamefont {Thompson}, \citenamefont {Aktulga}, \citenamefont {Berger}, \citenamefont {Bolintineanu}, \citenamefont {Brown}, \citenamefont {Crozier}, \citenamefont {in~'t Veld}, \citenamefont {Kohlmeyer}, \citenamefont {Moore}, \citenamefont {Nguyen}, \citenamefont {Shan}, \citenamefont {Stevens}, \citenamefont {Tranchida}, \citenamefont {Trott},\ and\ \citenamefont {Plimpton}}]{Thompson2022LAMMPSScales}%
  \BibitemOpen
  \bibfield  {author} {\bibinfo {author} {\bibfnamefont {A.~P.}\ \bibnamefont {Thompson}}, \bibinfo {author} {\bibfnamefont {H.~M.}\ \bibnamefont {Aktulga}}, \bibinfo {author} {\bibfnamefont {R.}~\bibnamefont {Berger}}, \bibinfo {author} {\bibfnamefont {D.~S.}\ \bibnamefont {Bolintineanu}}, \bibinfo {author} {\bibfnamefont {W.~M.}\ \bibnamefont {Brown}}, \bibinfo {author} {\bibfnamefont {P.~S.}\ \bibnamefont {Crozier}}, \bibinfo {author} {\bibfnamefont {P.~J.}\ \bibnamefont {in~'t Veld}}, \bibinfo {author} {\bibfnamefont {A.}~\bibnamefont {Kohlmeyer}}, \bibinfo {author} {\bibfnamefont {S.~G.}\ \bibnamefont {Moore}}, \bibinfo {author} {\bibfnamefont {T.~D.}\ \bibnamefont {Nguyen}}, \bibinfo {author} {\bibfnamefont {R.}~\bibnamefont {Shan}}, \bibinfo {author} {\bibfnamefont {M.~J.}\ \bibnamefont {Stevens}}, \bibinfo {author} {\bibfnamefont {J.}~\bibnamefont {Tranchida}}, \bibinfo {author} {\bibfnamefont {C.}~\bibnamefont {Trott}},\ and\ \bibinfo {author} {\bibfnamefont {S.~J.}\ \bibnamefont {Plimpton}},\ }\href
  {https://doi.org/10.1016/J.CPC.2021.108171} {\bibfield  {journal} {\bibinfo  {journal} {Comput. Phys. Commun.}\ }\textbf {\bibinfo {volume} {271}},\ \bibinfo {pages} {108171} (\bibinfo {year} {2022})}\BibitemShut {NoStop}%
\bibitem [{\citenamefont {Tribello}\ \emph {et~al.}(2014)\citenamefont {Tribello}, \citenamefont {Bonomi}, \citenamefont {Branduardi}, \citenamefont {Camilloni},\ and\ \citenamefont {Bussi}}]{Tribello2014PLUMEDBird}%
  \BibitemOpen
  \bibfield  {author} {\bibinfo {author} {\bibfnamefont {G.~A.}\ \bibnamefont {Tribello}}, \bibinfo {author} {\bibfnamefont {M.}~\bibnamefont {Bonomi}}, \bibinfo {author} {\bibfnamefont {D.}~\bibnamefont {Branduardi}}, \bibinfo {author} {\bibfnamefont {C.}~\bibnamefont {Camilloni}},\ and\ \bibinfo {author} {\bibfnamefont {G.}~\bibnamefont {Bussi}},\ }\href {https://doi.org/10.1016/J.CPC.2013.09.018} {\bibfield  {journal} {\bibinfo  {journal} {Comput. Phys. Commun.}\ }\textbf {\bibinfo {volume} {185}},\ \bibinfo {pages} {604} (\bibinfo {year} {2014})}\BibitemShut {NoStop}%
\bibitem [{\citenamefont {Bussi}\ \emph {et~al.}(2007)\citenamefont {Bussi}, \citenamefont {Donadio},\ and\ \citenamefont {Parrinello}}]{Bussi2007CanonicalRescaling}%
  \BibitemOpen
  \bibfield  {author} {\bibinfo {author} {\bibfnamefont {G.}~\bibnamefont {Bussi}}, \bibinfo {author} {\bibfnamefont {D.}~\bibnamefont {Donadio}},\ and\ \bibinfo {author} {\bibfnamefont {M.}~\bibnamefont {Parrinello}},\ }\href {https://doi.org/10.1063/1.2408420/186581} {\bibfield  {journal} {\bibinfo  {journal} {J. Chem. Phys.}\ }\textbf {\bibinfo {volume} {126}},\ \bibinfo {pages} {14101} (\bibinfo {year} {2007})}\BibitemShut {NoStop}%
\bibitem [{\citenamefont {Ray}\ \emph {et~al.}(2022)\citenamefont {Ray}, \citenamefont {Ansari}, \citenamefont {Rizzi}, \citenamefont {Invernizzi},\ and\ \citenamefont {Parrinello}}]{ray2022rare}%
  \BibitemOpen
  \bibfield  {author} {\bibinfo {author} {\bibfnamefont {D.}~\bibnamefont {Ray}}, \bibinfo {author} {\bibfnamefont {N.}~\bibnamefont {Ansari}}, \bibinfo {author} {\bibfnamefont {V.}~\bibnamefont {Rizzi}}, \bibinfo {author} {\bibfnamefont {M.}~\bibnamefont {Invernizzi}},\ and\ \bibinfo {author} {\bibfnamefont {M.}~\bibnamefont {Parrinello}},\ }\href {https://doi.org/10.1021/acs.jctc.2c00806} {\bibfield  {journal} {\bibinfo  {journal} {J. Chem. Theory Comput.}\ }\textbf {\bibinfo {volume} {18}},\ \bibinfo {pages} {6500} (\bibinfo {year} {2022})}\BibitemShut {NoStop}%
\end{thebibliography}%

\end{document}

\title{Supporting Information: Weighted Active Space Protocol for Multireference Machine-Learned Potentials} % for ab-inition dynamics simulations for catalysis

\author{Aniruddha Seal}\thanks{These authors contributed equally to this work.}
\affiliation{Department of Chemistry and Chicago Center for Theoretical Chemistry, University of Chicago, Chicago, IL 60637, USA}

\author{Simone Perego}\thanks{These authors contributed equally to this work.}
\affiliation{Atomistic Simulations, Italian Institute of Technology, 16156 Genova, Italy}

\author{Matthew R. Hennefarth}
\affiliation{Department of Chemistry and Chicago Center for Theoretical Chemistry, University of Chicago, Chicago, IL 60637, USA}

\author{Umberto Raucci}
\affiliation{Atomistic Simulations, Italian Institute of Technology, 16156 Genova, Italy}

\author{Luigi Bonati}
\affiliation{Atomistic Simulations, Italian Institute of Technology, 16156 Genova, Italy}

\author{Andrew L. Ferguson} 
\affiliation{Department of Chemistry and Chicago Center for Theoretical Chemistry, University of Chicago, Chicago, IL 60637, USA}
\affiliation{Pritzker School of Molecular Engineering, University of Chicago, Chicago, IL 60637, USA}

\author{Michele Parrinello} \email[corresponding author: ]{michele.parrinello@iit.it}
\affiliation{Atomistic Simulations, Italian Institute of Technology, 16156 Genova, Italy}

\author{Laura Gagliardi} \email[corresponding author: ]{lgagliardi@uchicago.edu} 
\affiliation{Department of Chemistry and Chicago Center for Theoretical Chemistry, University of Chicago, Chicago, IL 60637, USA}
\affiliation{Pritzker School of Molecular Engineering, University of Chicago, Chicago, IL 60637, USA}

\date{\today}

\maketitle

\tableofcontents

%\vspace{5cm}

\newpage

\section{Active space: Orbitals and NVE consistency check}

\begin{figure}[h!]
    \includegraphics[width=0.9\columnwidth]{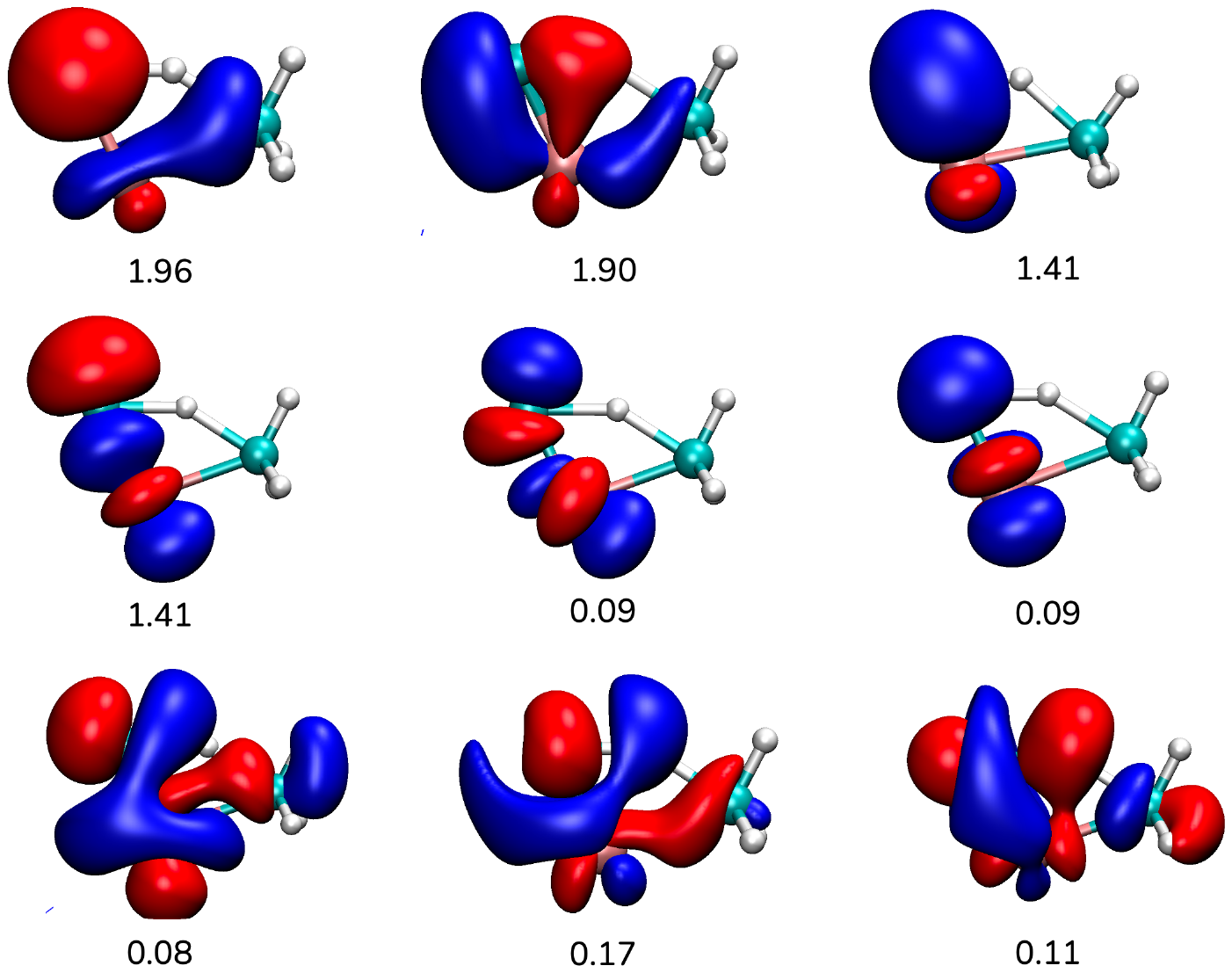}
    \caption{(7, 9) active space natural orbitals and their occupation numbers at the transition state of the reaction.}
    \label{fig:active_space}
\end{figure}

\begin{figure}[h!]
    \includegraphics[width=0.9\linewidth]{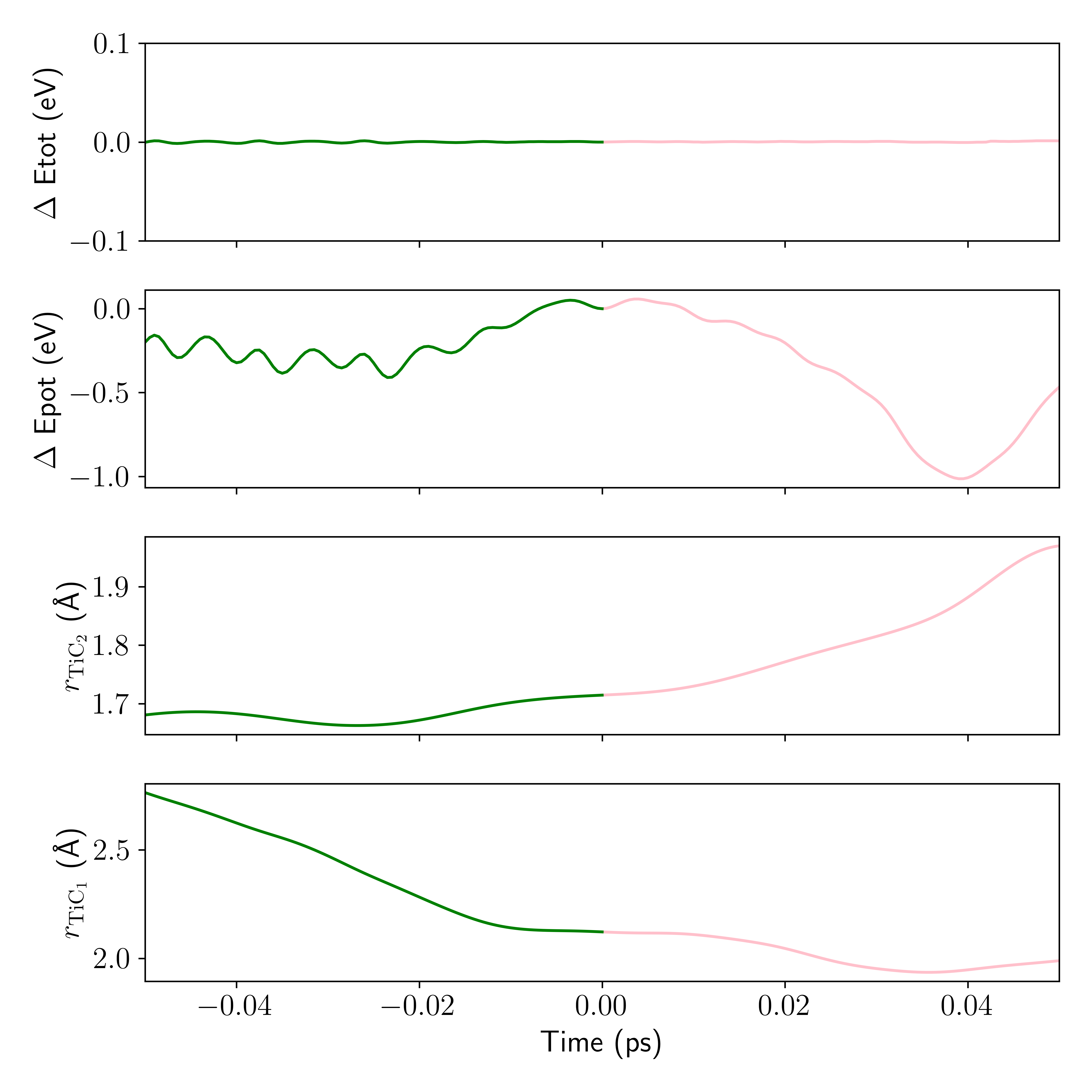}
    \caption{Relative change in (a) total energy and (b) potential energy along two NVE trajectories initiated from the optimized \textbf{TS} structure of the reaction. The trajectories were propagated using SA(2)-tPBE(7,9)/def2-TZVP, with an initial velocity sampled at \qty{300}{\kelvin} in opposite directions. The progress of the trajectories are tracked using the \ce{Ti-C} distances, (c) distance of \ce{Ti} from the \ce{C} on \ce{TiC+} and (d) distance of \ce{Ti} from the bonding methane carbon.}
    \label{fig:nve_e_con}
\end{figure}

To validate the flexibility of (7,9) active space shown in \Cref{fig:active_space}  used for the complete \textbf{R} $\rightarrow$ \textbf{P} transition, we assess its ability to consistently describe both the reactants (\ce{TiC+} and \ce{CH4}) and the product (\ce{HC-Ti-CH3}). We perform a molecular dynamics simulation initiated from the optimized \textbf{TS} structure, integrating the trajectory using a velocity-Verlet scheme in the microcanonical (NVE) ensemble. Initial velocities are sampled from the Maxwell–Boltzmann distribution at 300 K. As shown in \cref{fig:nve_e_con}, forward integration over 100 steps leads to the formation of products, while integration with reversed velocities recovers the reactants as can be monitored using the two \ce{Ti-C} distances. In both cases, the total energy remains stable without unphysical discontinuities, indicating that the active space remains consistent throughout the simulation. This demonstrates that the chosen active space is sufficiently flexible to accurately describe all relevant geometries along the reaction pathway.

\section{Parametrizing $\delta$ in WASP}

\begin{figure*}
  \includegraphics[width=0.8\linewidth]{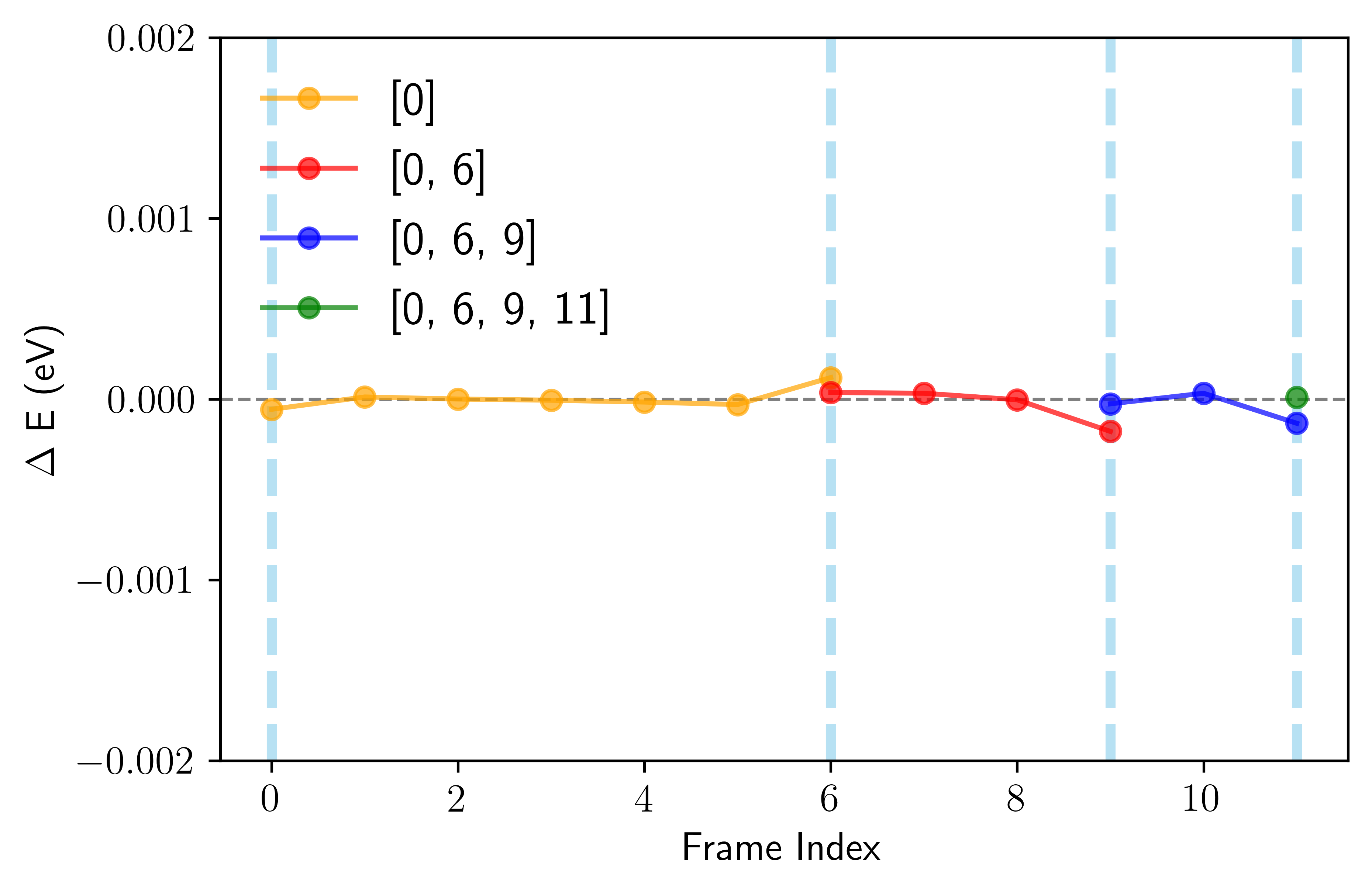}
  \caption{\label{fig:chk} Selection of WASP checkpoints along the reaction pathway. Plotted are the MC-PDFT energy deviations computed using WASP active-space guesses for each geometry following a given checkpoint, up to the point where the energy threshold is breached. Reference MC-PDFT energies were obtained by initializing each calculation with a fully converged CASSCF wave function from the previous geometry.}
\end{figure*}

A library of checkpoint files ($\Bqty{\vb{G}_\alpha}$, $\Bqty{\vb{C}_\alpha}$) contains each molecular geometry $\vb{G}_\alpha$ and the corresponding MO coefficient matrix $\vb{C}_\alpha$. In order to define a library of checkpoints for the set of geometries along a reaction pathway to test and parametrize WASP, we employed the following procedure. Initially, the starting geometry and wave function pair ($\vb{G}_0$, $\vb{C}_0$) computed based on the active space shown in \cref{fig:active_space} was added to the library. For subsequent geometries, the WASP scheme (with $\delta$ set to $\infty$, that is with no neighborhood filter) was used to compute starting guesses from the existing library of checkpoints. If the computed energy did not agree within the tolerance of \qty{1e-7}{\hartrees}, the corresponding geometry-wave function pair was added to the library. \Cref{fig:chk} shows that the procedure ultimately resulted in just four geometries out of twelve being retained in the library.

\begin{figure*}
  \includegraphics[width=\linewidth]{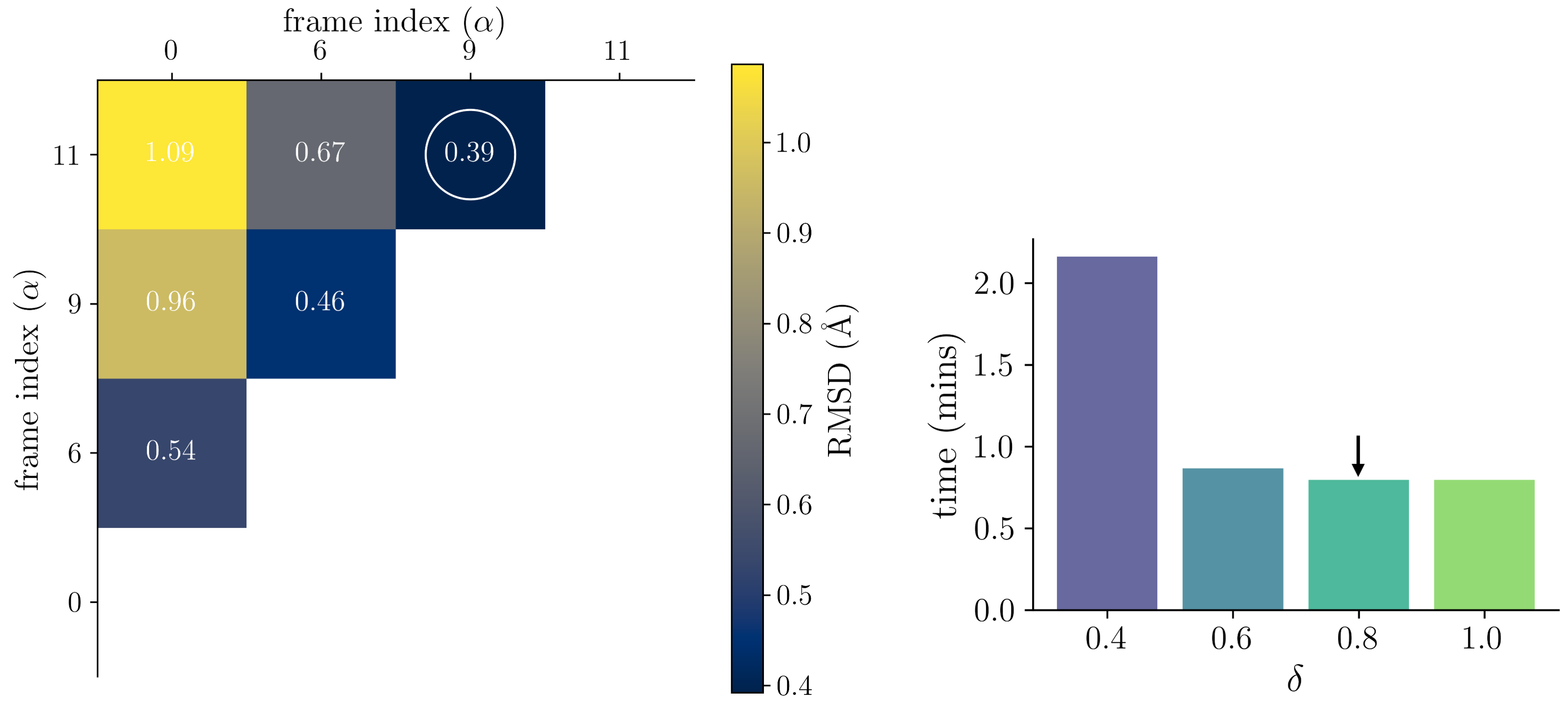}
  \caption{\label{fig:delta} Parameterizing the neighbor-cutoff parameter $\delta$ in WASP. (Left) Root-mean-squared-deviation (RMSD) computed for each of the checkpoint geometries with respect to one-another, used to set a lower bound. (Right) Average wall time per geometry for MC-PDFT calculations seeded with WASP active-space selection at various $\delta$ values. Note that the energy differences with all the tested $\delta$ values below the threshold compared to the reference.}
\end{figure*}

To determine an appropriate value for the cutoff parameter $\delta$, we analyzed the pairwise RMSD between the 4 checkpoint geometries retained by the WASP protocol shown in \cref{fig:delta}. The smallest RMSD value observed was \qty{0.4}{\angstrom}, which provides a natural lower bound for $\delta$. Because choosing a $\delta$ below this would result in no neighboring geometry being considered close enough to contribute a wave function guess. To ensure that each geometry could leverage neighbor information effectively, we tested $\delta$ values from \qty{0.4} to \qty{1}{\angstrom}. We found that a value of \qty{0.8}{\angstrom} strikes the best balance in capturing neighbor contributions, leading to the lowest overall wall time (\cref{fig:delta}).

\section{Active learning exploration statistics}

\begin{figure}[H]
    \centering
    \includegraphics[width=0.5\linewidth]{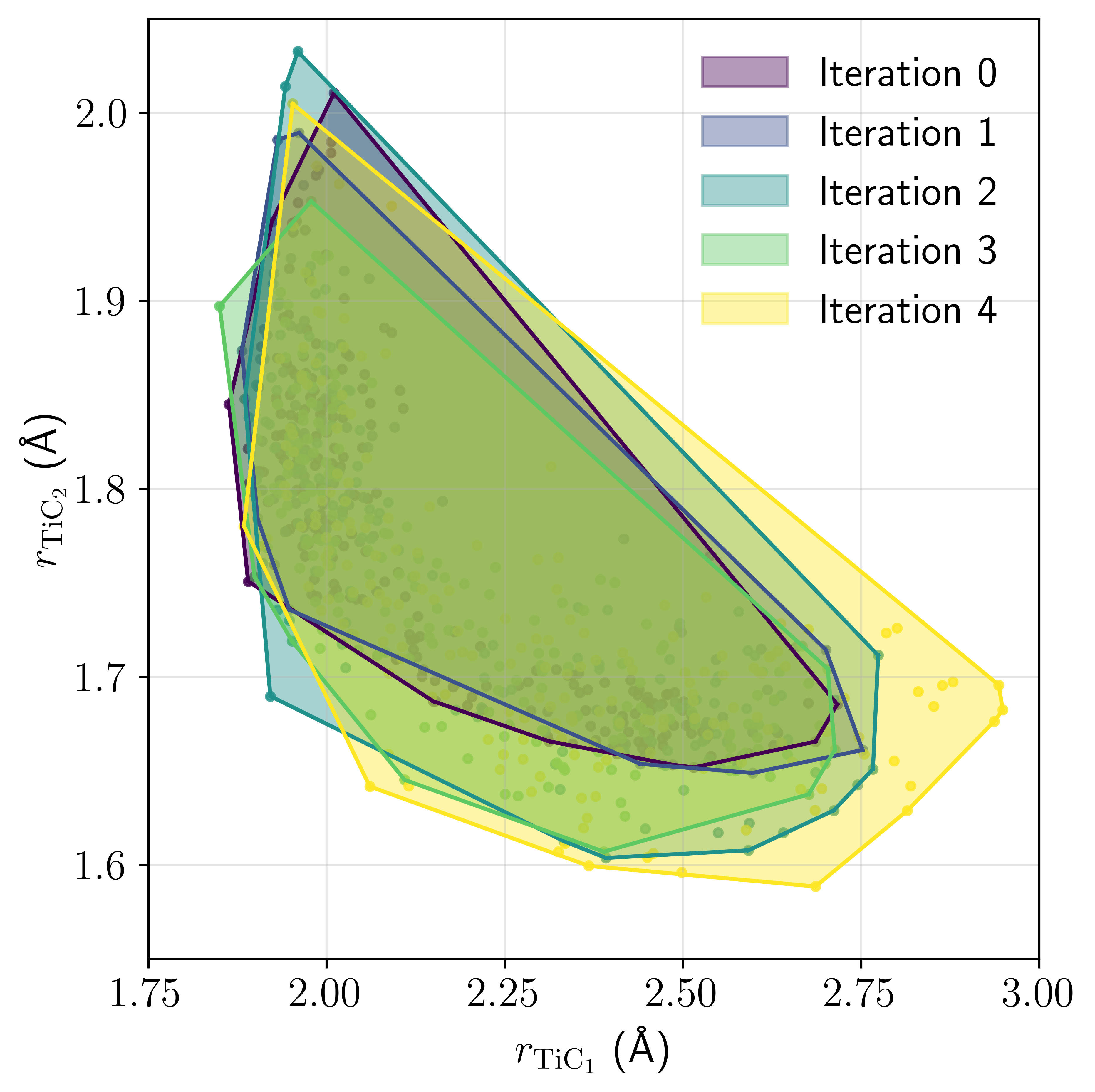}
    \caption{Convex hull illustrating the exploration of the configurational space across the progressive active learning rounds computed with the data is projected onto the plane defined by the two \ce{Ti-C} bond distances.}
    \label{fig:convex_hull}
\end{figure}

To visualize the progression of exploration during each active learning cycle described in the main text, a convex hull was computed with \texttt{scipy.spatial} using the data collected in each round of active learning. \Cref{fig:convex_hull} shows how the sampling extended into different regions of the configurational space in progressive active learning rounds.

\begin{figure}[H]
    \centering
    \includegraphics[width=0.5\linewidth]{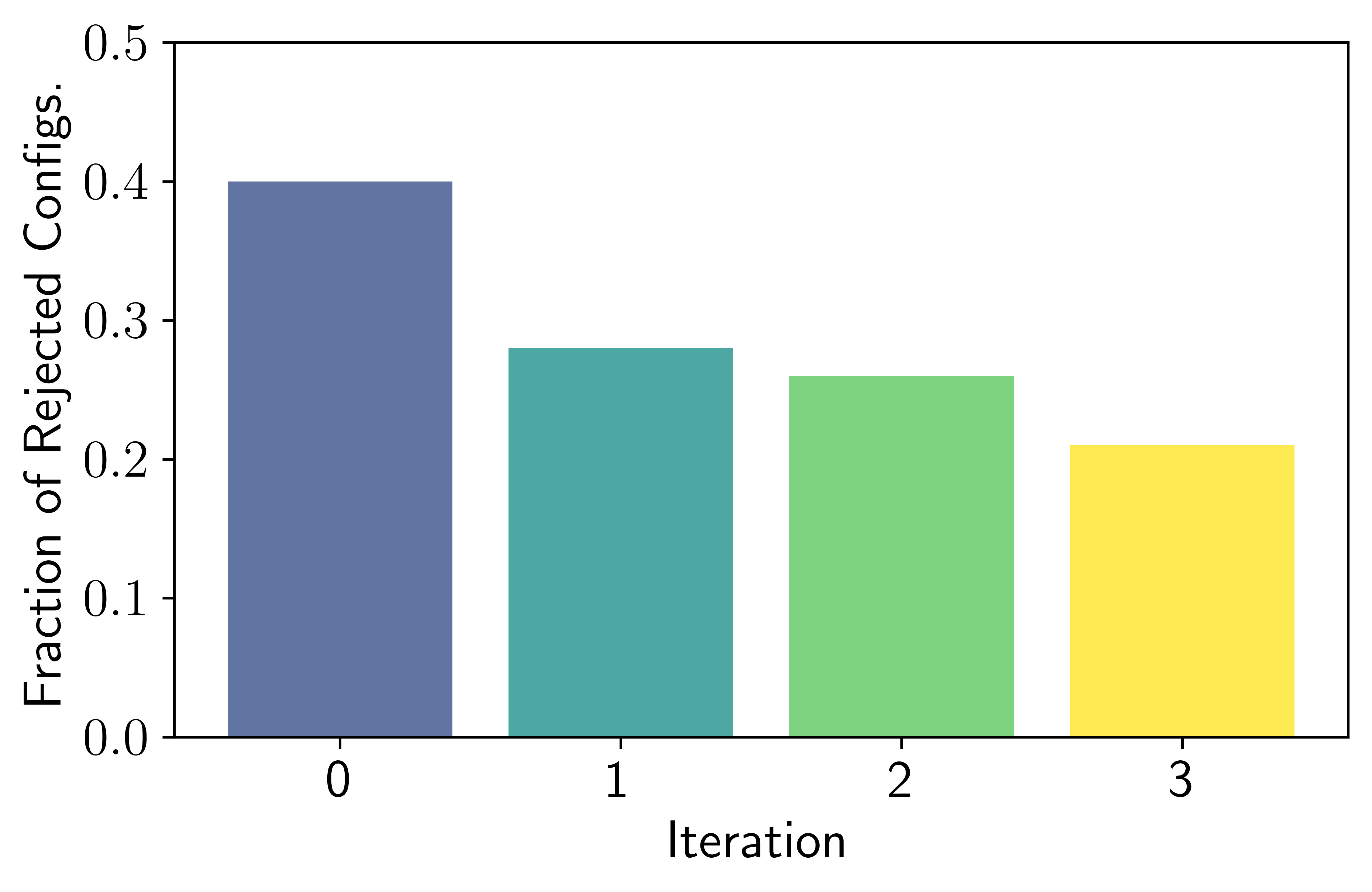}
    \caption{Fraction of rejected configurations for each active learning round}
    \label{fig:al_rej}
\end{figure}

In each active‐learning round, we define the fraction of rejected configurations as one minus the ratio of configurations retained after the incremental training-set update to those retained immediately following the DEAL step. \Cref{fig:al_rej} illustrates the consistent decrease in rejected configurations over successive active-learning rounds, indicating that more of the relevant configurational space is being effectively covered, thus supporting the robustness of the dual-filter strategy in the active-learning protocol.

\section{Additional results from MLP-based molecular dynamic simulations}

\begin{figure}[h!]
    \includegraphics[width=0.6\linewidth]{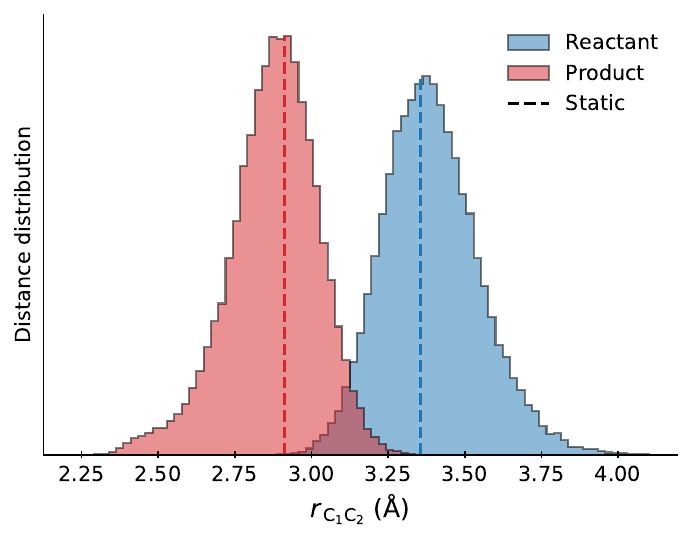}
    \caption{Dynamics analysis using a MLP trained on MC-PDFT data. Distribution of \ce{C-C} distance from equilibrium molecular dynamics simulations at \qty{300}{\kelvin} at the reactant (blue) and product (red) geometries. The static values are obtained from the \qty{0}{\kelvin} optimized reaction pathway reported in \cref{fig:NEB}.}
    \label{fig:equilibrium-dcc}
\end{figure}

\begin{figure}[h!]
    \includegraphics[width=0.6\linewidth]{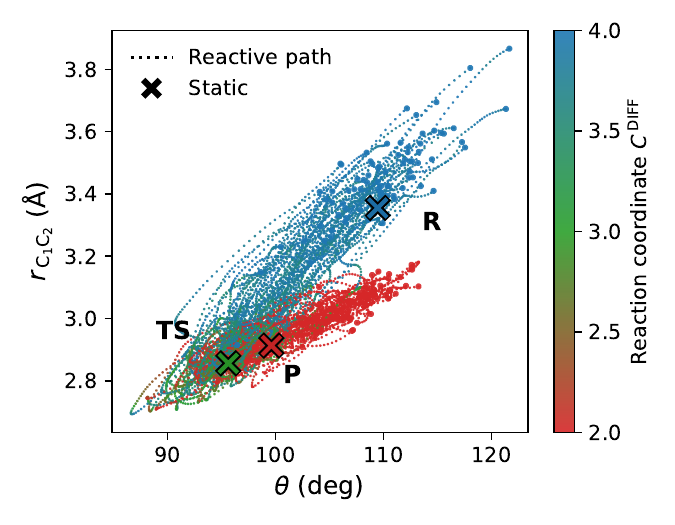}
    \caption{Dynamics analysis using a MLP trained on MC-PDFT data. Projection of reactive OPES flooding trajectories onto the distance between the two carbon atoms ($r_{\ce{C1C2}}$) and their angle ($\theta$), and colored by the difference between the two coordination number between carbon and hidorgen atoms ($C^{\text{DIFF}}$). 
    The static values are obtained from the \qty{0}{\kelvin} optimized reaction pathway reported in \cref{fig:NEB}.
    }
    \label{fig:path-dcc-angle}
\end{figure}

\section{Static reaction pathway with final MLP}

\begin{figure}[h!]
    \includegraphics[width=0.7\linewidth]{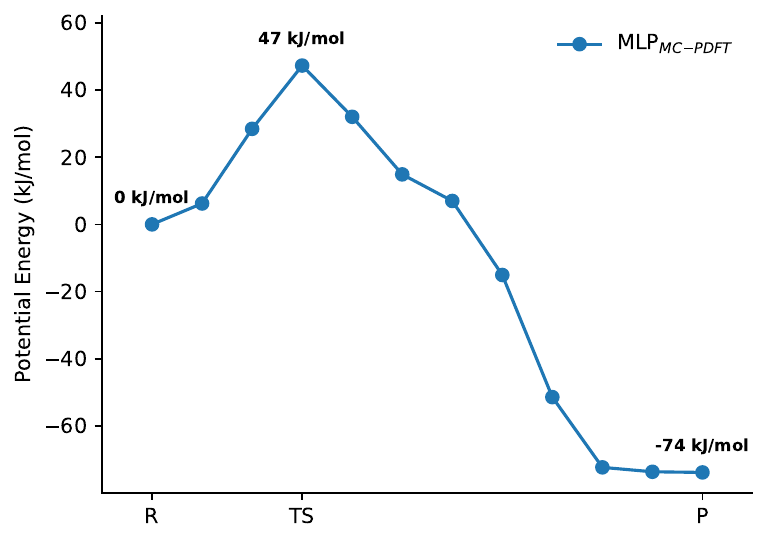}
    \caption{Minimum energy path computed via CI-NEB using the MACE MLP trained on MC-PDFT data. The potential energy values at key image -- the reactant (R), transition state (TS), and product (P) -- are annotated.
    }
    \label{fig:NEB}
\end{figure}

In \cref{fig:NEB}, we report the static reaction pathway using the MACE-based MLP trained on MC-PDFT data. Reactant and product configurations were obtained by geometry optimization using the FIRE algorithm \cite{bitzek2006structural} with a force convergence criterion of \qty{0.01}{\eV\per\angstrom}. The reactive path was computed using 12 images, using the climbing image nudge elastic band (CI-NEB) \cite{henkelman2000climbing} method, with a spring constant of \qty{0.5}{\eV\per\angstrom^{2}}. The NEB relaxation used the same FIRE optimizer with a force threshold of \qty{0.04}{\eV\per\angstrom}. 

\section{Estimate rates from OPES-flooding simulations}

To compute the rate constant for the transition between two metastable states, we employed the OPES-flooding method (see main text methods and materials). A series of 100 independent OPES-flooding simulations was performed, each initiated from the reactant basin. For each trajectory, we extracted the first passage time (FPT)—defined as the reweighted time required for the system to reach the product basin, as determined by a threshold in the CV space. 

\begin{figure}[H]
    \centering
    \includegraphics[width=0.6\linewidth]{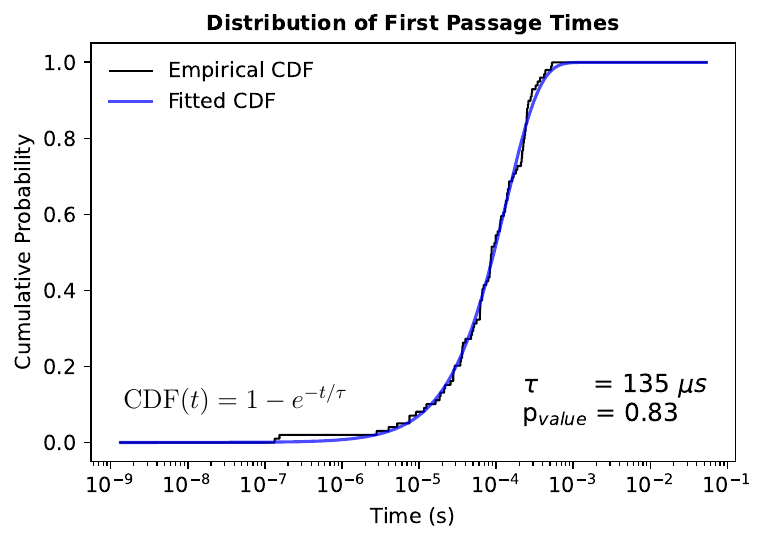}
    \caption{ Empirical cumulative distribution function (CDF) of the reweighted first passage times (black line) fitted with a single-exponential model (blue line). The fitted timescale is $\tau= \qty{135}{\micro\second}$, corresponding to a rate constant of \qty{5.1E3}{\second^{-1}}, and the p-value of the Kolmogorov–Smirnov test is reported in the figure.
}
    \label{fig:rates}
\end{figure}

To estimate the rate constant, we collected FPTs across all trajectories and constructed their empirical cumulative distribution function (CDF). The CDF was fitted to a single-exponential model $\mathrm{CDF}(t) = 1 - \exp(-t/\tau)$, where $\tau$ is the characteristic timescale of the transition (\cref{fig:rates}). The rate constant was then obtained as $k = 1/(\tau \mu)$, where $\mu$ is the mean first passage time. The quality of the exponential fit was assessed via the Kolmogorov–Smirnov test.

\section{Optimized Geometries}

The coordinates (reported in \textup{\AA}) and ground state energies of the geometries used in Figure 1 of the main text.

\begin{table}[H]
    \centering
    \caption{\textbf{R} geometry optimized at SA(2)-tPBE(7,9)/def2TZVP; E$_{\text{SA-CASSCF}}$ = -926.2366807 hartrees, E$_{\text{MC-PDFT}}$ = -927.2759341 hartrees}
    \begin{tabular}{lrrr}
    Ti & 9.99837 & 10.00085 & 10.00677 \\
C & 10.87608 & 12.23382 & 9.98688 \\
C & 11.27225 & 8.91449 & 9.96612 \\
H & 11.56416 & 11.34482 & 9.97334 \\
H & 11.60664 & 13.05078 & 9.95580 \\
H & 10.24435 & 12.35258 & 9.09092 \\
H & 10.30075 & 12.36692 & 10.91755 \\
    \end{tabular}
\end{table}

\begin{table}[H]
    \centering
    \caption{\textbf{P} geometry optimized at SA(2)-tPBE(7,9)/def2TZVP; E$_{\text{SA-CASSCF}}$ =-926.1801277 hartrees, E$_{\text{MC-PDFT}}$ = -927.2555934 hartrees}
    \begin{tabular}{lrrr}
    Ti & 9.99031 & 9.99474 & 9.99428 \\
C & 10.41613 & 11.90114 & 10.33274 \\
C & 11.47929 & 9.23715 & 10.74002 \\
H & 12.40013 & 8.86276 & 11.20681 \\
H & 11.34133 & 12.22275 & 9.83574 \\
H & 9.54236 & 12.41522 & 9.88369 \\
H & 10.47862 & 12.12888 & 11.40575 \\
    \end{tabular}
\end{table}

\begin{table}[H]
    \centering
    \caption{\textbf{TS} geometry from \citet{geng_intrinsic_2019}; E$_{\text{SA-CASSCF}}$ = -926.2313930 hartrees, E$_{\text{MC-PDFT}}$ = -927.3040846 hartrees}
    \begin{tabular}{lrrr}
    Ti & 0.42874 & 0.70325 & -0.02040 \\
C & 1.18856 & 2.77001 & -0.00429 \\
C & 2.00825 & -0.05807 & 0.02005 \\
H & 2.01663 & 1.37869 & 0.01136 \\
H & 2.16803 & 3.21948 & -0.02188 \\
H & 0.63867 & 3.10323 & -0.88498 \\
H & 0.67327 & 3.09352 & 0.90013 \\
    \end{tabular}
\end{table}

\bibliography{ref}